\documentclass[10pt,journal,compsoc]{IEEEtran}

\ifCLASSOPTIONcompsoc
  \usepackage[nocompress]{cite}
\else
  \usepackage{cite}
\fi

\ifCLASSINFOpdf
\else
\fi
\usepackage{graphicx}
\usepackage{soul}
\usepackage{subfig}
\usepackage[fleqn]{amsmath}
\usepackage{relsize}
\usepackage{algorithm}
\usepackage[noend]{algpseudocode}
\usepackage{varwidth}
\usepackage{ragged2e}
\usepackage{booktabs} % For formal tables
\usepackage[percent]{overpic}
\usepackage{multirow}
\usepackage{bm}
\usepackage{enumitem}
\usepackage[normalem]{ulem}
\usepackage{tikz}
\soulregister\cite7
\soulregister\ref7
\soulregister\pageref7

\begin{document}
\title{An Efficient Hybrid I/O Caching Architecture Using Heterogeneous SSDs}

\newcommand{\textoverline}[1]{$\overline{\mbox{#1}}$}
\newcommand{\specialcell}[2][c]{%
	\begin{tabular}[#1]{@{}l@{}}#2\end{tabular}}

\author{Reza Salkhordeh,
        Mostafa Hadizadeh, and~Hossein~Asadi% <-this % stops a space
\IEEEcompsocitemizethanks{\IEEEcompsocthanksitem Reza Salkhordeh, Mostafa Hadizadeh, and Hossein Asadi (corresponding author) are with the Department of Computer Engineering, Sharif University of Technology, Emails: salkhordeh@ce.sharif.edu, mhadizadeh@ce.sharif.edu, and asadi@sharif.edu .\protect\\
}% <-this % stops a space
}

\IEEEtitleabstractindextext{%
	{\justify
\begin{abstract}
Storage subsystem is considered as the performance bottleneck of computer systems in data-intensive applications.
\emph{Solid-State Drives} (SSDs) are emerging storage devices which unlike \emph{Hard Disk Drives} (HDDs), do not have mechanical parts and therefore, have superior performance compared to HDDs.
Due to the high cost of SSDs, \emph{entirely} replacing HDDs with SSDs is not economically justified.
Additionally, SSDs can {endure} a limited number of writes before failing.
To mitigate the shortcomings of SSDs while taking advantage of their high performance, SSD caching is practiced in both academia and industry.
Previously proposed caching architectures have \emph{only} focused on either performance or endurance and neglected to address \emph{both} parameters in suggested architectures.
Moreover, the cost, reliability, and power consumption of such architectures is not evaluated.
This paper proposes a hybrid I/O caching architecture that while offers higher performance than previous studies, it also improves power consumption with a similar budget.
The proposed architecture uses DRAM, \emph{Read-Optimized SSD} (RO-SSD), and \emph{Write-Optimized SSD} (WO-SSD) in a three-level cache hierarchy and tries to efficiently redirect read requests to either DRAM or RO-SSD while sending writes to WO-SSD.
To provide high reliability, dirty pages are written to at least two devices which removes any single point of failure.
The power consumption is also managed by reducing the number of accesses issued to SSDs.
The proposed architecture reconfigures itself between performance- and endurance-optimized policies based on the workload characteristics to maintain an effective tradeoff between performance and endurance.
We have implemented the proposed architecture on a server equipped with industrial SSDs and HDDs.
The experimental results show that as compared to state-of-the-art studies, the 
proposed architecture improves performance and power consumption by an average 
of {8\% and 28\%,} respectively, and reduces the cost by 5\% while 
{increasing the endurance cost by 4.7\%} and negligible reliability penalty.

%{Therefore, for} evaluating a SSD-based cache architecture, its performance, reliability, cost, and endurance should be considered simultaneously.
%Moreover, previously proposed caching architectures have \emph{only} focused on either performance or endurance and neglected to address \emph{both} parameters in suggested architectures.
%This paper proposes a hybrid I/O cache architecture that while offers higher performance than previous studies, it also improves power consumption, reliability, and endurance with a similar budget.
%The proposed architecture uses DRAM, \emph{Read-Optimized SSD} (RO-SSD), and \emph{Write-Optimized SSD} (WO-SSD) in a three-level cache hierarchy and tries to efficiently redirect read requests to either DRAM or RO-SSD while sending writes to WO-SSD.
%To provide high reliability, dirty pages are written to at least two devices which removes any single point of failure.
%Power consumption and endurance are improved by reducing the number of writes issues to SSDs.
%Experimental results show that the proposed architecture improves performance, power consumption, and endurance on average by 31\%, 11\%, and 57\%, respectively while having no reliability penalty with only 15\% cost overhead.
\end{abstract}}

}

\maketitle

% is not selected <OR> if conference mode is selected - because compsoc
% conference papers position the abstract like regular (non-compsoc)
\IEEEdisplaynontitleabstractindextext

% \fi
\IEEEpeerreviewmaketitle

\section{Introduction}
\emph{Hard Disk Drives} (HDDs) are traditional storage devices that are commonly used in storage systems due to their low cost and high capacity.
The performance gap between HDDs and other components of computer systems has significantly increased in the recent years.
This is due to HDDs have mechanical parts which puts an upper limit on their performance.
To compensate the low performance of HDDs, storage system designers proposed several hybrid architectures consists of HDDs and faster storage devices such as \emph{Solid-State Drives} (SSDs).

\begin{figure}[t]
	\centering
	\includegraphics[scale=0.65]{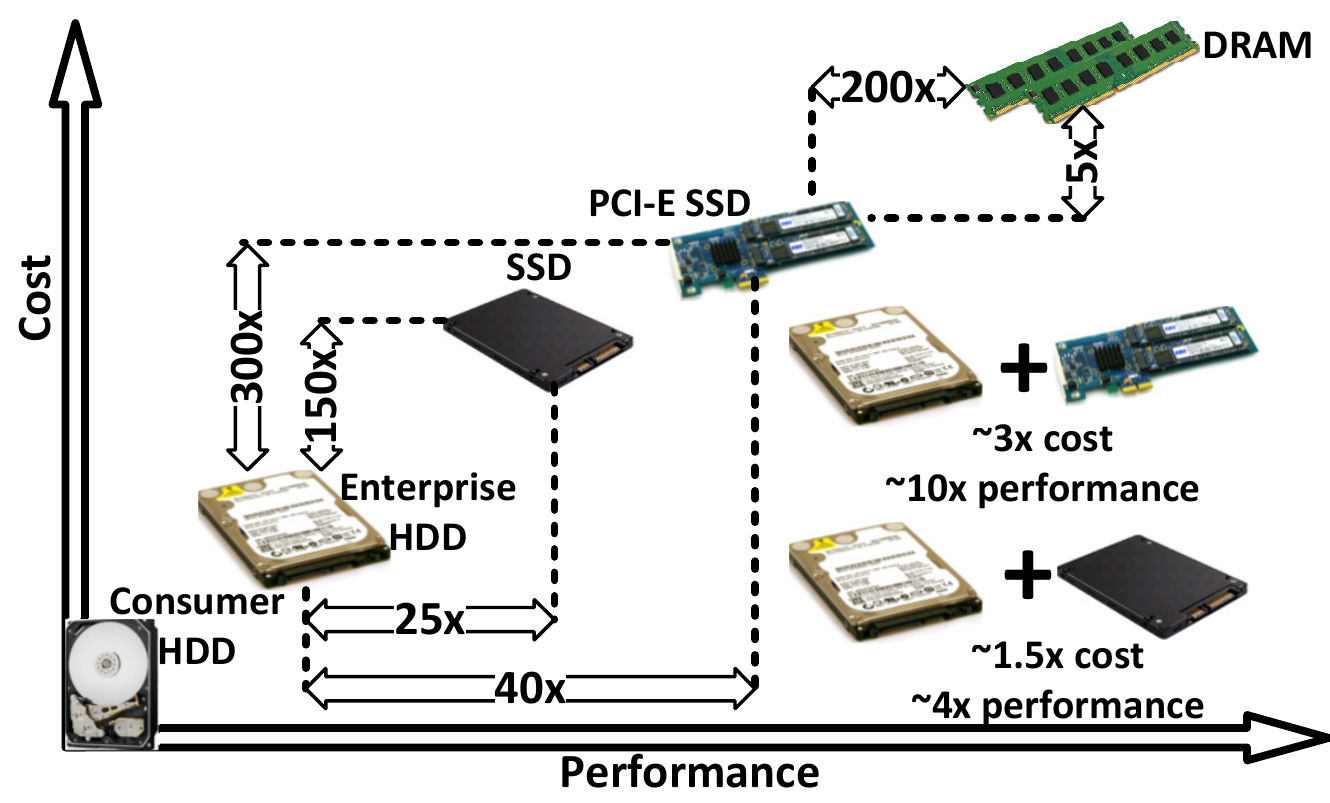}
	\caption{Storage Devices Characteristics}
	\label{fig:devices}
	\vspace{-0.6cm}
\end{figure}

SSDs are non-mechanical storage devices that offer higher performance in random workloads and asymmetric read/write performance as compared to HDDs.
SSD manufacturers design and produce several types of SSDs with different performance and cost levels to match a wide range of user requirements.
The relative performance and costs of SSDs compared to HDDs and \emph{Dynamic Random Access Memory} (DRAM) is shown in Fig. \ref{fig:devices}.
Due to the relatively very high price of SSDs, replacing the entire disk array in data storage systems with SSDs is not practical in \emph{Big Data} era.
In addition, SSDs have restricted lifetime due to the limited number of reliable writes which can be committed to SSDs.
The power outage can also cause data loss in SSDs as reported in \cite{ahmadian:date}.
Although SSDs have such shortcomings, they have received a significant attention from both academic and industry and many architectures for I/O stack based on SSDs have been proposed in recent years.
%One of the main approaches for efficiently employing such expensive devices, is using them for storing performance-critical data which makes the paid cost justifiable.

One promising application of SSDs emerged in recent years is to alleviate low performance of HDDs with minimal cost overhead by using SSDs as a caching layer for HDD-based storage systems.
The main focus of previous studies in caching architecture is on improving performance and/or endurance.
Three main approaches have been proposed in previous studies to this end: a) prioritizing request types such as filesystem metadata, random, and read requests, b) optimizing baseline algorithms, and c) modifying the traditional single-level cache.
As shown in Fig. \ref{fig:devices}, caching architectures offer various performance levels with significantly different costs depending on their choice of SSDs.
Previous studies neglected to consider the effect of choosing different SSDs on the performance.
Additionally, they are mostly focused on the performance, while other major system metrics such as power consumption, endurance, and reliability also need to be considered in the caching architectures.
To our knowledge, none of previous studies considered all the mentioned parameters, simultaneously.

This paper proposes a \emph{Three-level I/O Cache Architecture} (TICA) which aims to improve the performance and power consumption of SSD-based I/O caching while having minimal impact on the endurance.
TICA employs \emph{Read-Optimized SSD} (RO-SSD), \emph{Write-Optimized SSD} (WO-SSD), and DRAM as \emph{three levels} of I/O cache.
Employing heterogeneous SSDs decreases the probability of correlated failure of SSDs since such SSDs either belong to different brands or have different internal data structures/algorithms.
To enhance the performance of both read and write requests, TICA is configured in the write-back mode such that it buffers the most frequent read and write intensive data pages in DRAM to reduce the number of writes committed to SSDs and to increase their lifespan.
In order to guarantee the reliability of write requests, all write requests are committed to both DRAM and WO-SSD before the write is acknowledged to the user.
Dirty data pages in DRAM are asynchronously flushed to RO-SSD to free up allocated DRAM space for future requests.

In order to efficiently optimize the proposed architecture for read- and write-intensive applications, we offer two cache policies where evicted data pages from DRAM can be either moved to SSDs or removed from the cache.
The first policy, called \emph{Write to Endurance Disk} (TICA-WED), improves performance since the next access to the data page will be supplied by SSDs instead of HDD.
The shortcoming of TICA-WED is reducing SSDs lifetime due to the extra writes for moving the data page from DRAM to SSD.
To alleviate such shortcoming, the second policy, called \emph{Endurance Friendly} (TICA-EF), can be employed.
In TICA-EF, performance is slightly decreased while the lifetime of SSDs is significantly extended.
To select between TICA-WED and TICA-EF, we propose a state-machine which analyzes the running workload and dynamically selects the most effective policy for TICA.
With such data flow, TICA improves performance and power consumption of I/O cache while having negligible endurance overhead and no cost and reliability impact.

To verify the efficiency of TICA, we have first extracted I/O traces from a server equipped with two Intel Xeon, 32GB memory, and 2x SSD 512GB.
I/O traces are extensively analyzed and characterized to help optimize parameters of TICA towards higher performance.
Experimental setup consists of a rackmount server equipped with a RO-SSD, a WO-SSD, and 128GB memory.
The benchmarking suites for experiments consist of over 15 traces from Microsoft research traces \cite{snia1}, HammerDB \cite{hammerDB}, and FileBench {\cite{tarasov2016filebench}}.
Experimental results show that despite reducing the cost by 5\%, as compared to 
the state-of-the-art architectures, TICA enhances performance and power 
consumption, on average, by {8\%} (and up to {45\%}), and by {28\%} (and 
up to {70\%}), respectively, while having only {4.7\%} endurance overhead 
and negligible reliability penalty.

To our knowledge, we make the following contributions:
\begin{itemize}
\item By carefully analyzing state-of-the-art SSDs available in the market and their characteristics, we select two types of SSDs to design a low-cost hybrid caching architecture capable of providing high performance in both read- and write-intensive applications.
\item We propose a three-level caching architecture, called TICA, employing DRAM, RO-SSD, and WO-SSD to improve performance and power consumption of storage systems while having negligible endurance penalty.
\item TICA reduces the correlated failure rate of SSDs in I/O caching architectures by using heterogeneous SSDs while the performance is not limited by the slower SSD, unlike traditional heterogeneous architectures.
\item To balance performance and endurance, we propose \emph{Endurance-Friendly} (TICA-EF) and \emph{Write to Endurance Disk} (TICA-WED) policies for TICA, where the first policy prioritizes endurance and the second policy tries to further improve performance.
\item We also propose a state-machine model to select one of TICA-EF or TICA-WED policies based on the workload characteristics.
Such model can identify the most effective policy for TICA with negligible overhead while running I/O intensive applications.
\item We have implemented TICA on a physical server equipped with enterprise SSDs and HDDs and conducted an extensive set of experiments to accurately evaluate TICA, considering all optimization and buffering in storage devices and \emph{Operating System} (OS).
\end{itemize}

The remainder of this paper is organized as follows.
Previous studies are discussed in Section \ref{sec:previous}.
The motivation for this work is presented in Section \ref{sec:motivation}.
Section \ref{sec:proposed} introduces the proposed caching architecture.
In Section \ref{sec:results}, the experimental setup and results are presented.
Finally, Sec. \ref{sec:conclusion} concludes the paper. 

\section{Previous Studies}
\label{sec:previous}
Previous studies in 	SSD-based I/O caching can be categorized into three groups: a) prioritizing various request types based on storage device characteristics, b) optimizing baseline eviction and promotion policies, and c) proposing multi-level caching architectures.
The first category tries to characterize performance of HDDs and SSDs.
Based on the characterization, request types which have higher performance gap between HDDs and SSDs are prioritized to be buffered.
A comprehensive study on the workload characteristics and request types is conducted in \cite{7155523}.
The different response time of SSDs on read and write requests is considered in \cite{5366764} to prioritize data pages.
The locality of data pages is employed in RPAC \cite{7255203} to improve both performance and endurance of caching architecture.
ReCA \cite{8265001} tries to characterize several requests and workload types and selects suitable data pages for caching.
Filesystem metadata is one of the primary request types which is shown to be very efficient for caching \cite{tiering,macss}.
OODT \cite{tiering} considers randomness and frequency of accesses to prioritize the data pages.
To reduce the migrations between HDD and SSD, \cite{8085094} considers the dirty state of the data pages in memory buffers.
ECI-Cache \cite{ahmadian:eci} prioritizes data pages based on the request type (read/write) in addition to the reuse distance.
The optimization of the previous studies in this category is mostly orthogonal to TICA and can be employed in the eviction/promotion policies of the SSDs in TICA.

The studies in the second category try to optimize the eviction policy of caching architectures.
To prevent cache pollution, \emph{Lazy Adaptive Replacement Cache} (LARC) \cite{Huang:2016:IFD:2888404.2737832} is suggested which promotes data pages to cache on the second access to the data page.
This technique, however, cannot perform in a timely fashion when workload is not stable.
mARC \cite{Santana:2015:AA:2813749.2813763} tries to select the more suitable option from ARC and LARC based on the workload characteristics.
In \cite{6558445}, various management policies based on ARC for DRAM-SSD caching architectures are compared.
A more general approach to prevent repetitive replacement of data pages in SSDs is suggested in \cite{Liang:2016:EQU:2928275.2928286} which provides buffered data pages a more chance to be accessed again and therefore stay in the cache.
{S-RAC \cite{Ni:2016:SSF:2928275.2928284} characterizes workloads into six groups.
Based on the benefit of buffering requests in each category, it decides which data pages are best suited for caching.
S-RAC tries to reduce the number of writes in SSD to improve its lifetime with minimal impact on the cache performance.}
H-ARC \cite{6855546} partitions the cache space into \emph{clean} and \emph{dirty} sections where each section is maintained by ARC algorithm.
D-ARC \cite{7327162} also tries to improve ARC by prioritizing the data pages based on the clean/dirty state. 
Me-CLOCK \cite{7307128} tries to reduce the memory overhead of SSD caching architectures by using bloom filter.
{RIPQ \cite{188468} suggests a \emph{segmented-LRU} caching algorithm, which aggregates small random writes and also places user data with the same priority close to each other.}
In WEC \cite{7035001}, write-efficient data pages are kept in cache for longer periods to reduce the writes due to the cache replacements. 
This category of previous studies is also orthogonal to TICA and such policies can be employed jointly with TICA to further improve performance and/or endurance.

Among previous studies that try to enhance performance of I/O caching by utilizing multi-level cache hierarchies,
LLAMA \cite{Levandoski:2013:LCS:2536206.2536215} employs a DRAM-SSD architecture for designing an \emph{Application Programming Interface} (API) suitable for database management systems.
{FASC \cite{10.1007/978-3-319-18120-2_10} suggests a DRAM-SSD buffer cache, which tries to reduce the cost of evictions from buffer cache as well as write overheads on the SSD.}
Employing exclusive DRAM-SSD caching is investigated in \cite{6558445} which shows the impact of such technique on improving SSD endurance.
In \cite{doi:10.1002/cpe.3530}, separate promotion/demotion policies for DRAM/SSD cache levels are replaced with a unified promotion/demotion policy to improve both performance and endurance.
uCache \cite{6831945} also employs a DRAM-SSD architecture and tries to reduce the number of writes in the SSD due to the read misses.
In case of a power loss, all dirty data pages in DRAM will be lost which significantly reduces the reliability of uCache.
Additionally, no redundant device is employed and both DRAM and SSD are single points of failure.
MDIS \cite{7425254} uses a combination of DRAM, SSD, and NVM to improve performance of I/O caching.
Although performance and energy consumption are improved in MDIS, the cost and reliability have not been taken into account.
{Graphene \cite{Liu:2017:GFI:3129633.3129659} suggests a DRAM-SSD 
architecture to improve performance of graph computing for large graphs.
SSD caching is also suggested in distributed and \emph{High\space 
Performance\space Computing} (HPC) environments 
\cite{7390309,Kakoulli:2017:ODF:3035918.3064023,8025645,7776880,Arteaga:2016:COF:2930583.2930610}.}

{Optimizing SSDs for key-value store is discussed in previous studies.
DIDACache \cite{Shen:2017:DDI:3129633.3129668} allows the key-value SSD cache to directly manage the internal SSD structure to improve both performance and endurance.
WiscKey \cite{194424} separates key and value storage in SSD to improve random lookups and database loading.}
{Deduplication and compression can also be employed} to extend the SSDs lifetime {\cite{Li:2016:CID:2930583.2930606,8249847,194420}}.
Modifying the existing interface between OS and SSDs is also suggested in previous studies to design efficient caching architectures {\cite{Saxena:2014:DPS:2661087.2629491,194456}.
In \cite{194456}, a new interface for SSDs is designed, which does not allow overwriting of data pages, to reduce the size of the required DRAM in SSD and also to improve performance.
F2FS \cite{188454} employs an append-only logging approach to reduce the need for overwriting data pages in SSDs.
KAML \cite{7920840} suggests a customized interface for SSDs for storing and retrieval of key-value data.
FStream \cite{210522} employs \emph{streamID} to hint} \emph{Flash Translation 
Layer} {(FTL) on lifetime of user data so that FTL places the 
data pages with the same lifetime on a physical block.}
Optimizing SSD caching architectures by leveraging information from SSDs internal data structures such as FTL is also suggested in previous studies \cite{7425233,7967100}.
{FLIN \cite{8416843} provides a fair scheduler for SSDs servicing multiple 
applications simultaneously.
A scheduler to maximize the efficiency of parallelism inside of the SSD is also proposed in \cite{Elyasi:2017:EIS:3037697.3037728}.
SHRD \cite{Kim:2017:SIS:3129633.3129658} tries to optimize the physical placement of data pages in SSD to reduce the FTL overheads on random write requests.
AGCR \cite{194422} characterizes the workload behavior and increases the 
program time of  read-intensive data pages in the flash chips so that their 
read time can be decreased.}
Such architectures require hardware modifications which is not in the scope of this paper.

In general, one of the main shortcomings of previous studies is neglecting to consider the difference between various SSD brands and models in terms of cost and read/write performance.
Many types of SSDs are optimized towards read operations while others are optimized to provide higher write performance.
In addition, the tradeoff between performance, power consumption, endurance, reliability, and cost has not been considered in previous works which is crucial for I/O caching architectures. 

\section{Motivation}
\label{sec:motivation}
In this section, we detail the three shortcomings of state-of-the-art caching architectures which motivates us to propose three-level caching architecture employing SSDs in addition to DRAM.
First, we show the diverse characteristics of the SSDs in the market and the performance impact of employing such SSDs as the caching layer for HDDs.
Second, we evaluate the write overhead of caching read misses in SSDs.
Finally, we investigate the performance of mirrored heterogeneous SSDs employed to overcome the correlated SSDs failure.

\begin{figure}
	\centering
	\includegraphics[scale=0.45]{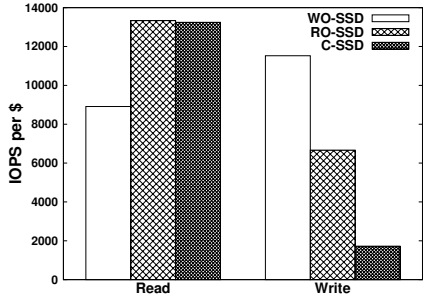}
	\caption{Performance per Cost for various SSDs}
	\label{fig:perfcost}
	%\vspace{-0.6cm}
\end{figure}

SSD manufacturers employ \emph{Single-Level Cell} (SLC), \emph{Multi-Level Cell} (MLC), or \emph{Three-Level Cell} (TLC) NAND chips in their products.
SLC SSDs have the highest performance and endurance at the cost of more than 2x of MLC SSDs.
The read performance of MLC SSDs, however, is comparable to the SLC SSDs due to the nature of the NAND flashes.
Table \ref{tab:cachingdevices} reports the performance and endurance of several types of SSDs.
Using high cost SSDs is not economically justifiable in several workload types.
Fig. \ref{fig:perfcost} shows the read and write IOPS per \$ for various SSDs.
In read-intensive workloads employing RO-SSD or \emph{Consumer-SSD} (C-SSD) results in higher performance per cost.
RO- or C-SSDs, however, fail to provide high performance per cost in write-intensive workloads.
This experiment reveals that high-cost and low-cost SSDs can be efficient in different workload types and using only one SSD type cannot provide suitable performance per cost in all workload types.

\begin{table}[t]
	\centering
	\caption{Power Consumption, Cost, Reliability, and Endurance of Storage Devices in the Cache Architectures
		\label{tab:cachingdevices}}  
	%       \vspace{-0.2cm}
	\begin{tabular}{|c|c|c|c|c|}
		\hline
		\textbf{Device} &  \textbf{MTTF (h)} & \textbf{\$/GB} & \textbf{Writes/GB} & \textbf{\shortstack{Read/Write/Idle\\Power (w)}} \\     \hline
		DRAM & 4M & 7.875 & $\infty$ & 4/4/4 \\     \hline
		C-SSD &  1.5M & 0.375   & 750 & 3.3/3.4/0.07 \\     \hline
		RO-SSD &  2M & 0.74   & 1,171 & 3.3/3.4/0.07\\     \hline
		WO-SSD & 2M & 0.842 & 6,416 & 2.4/3.1/1.3 \\     \hline
	\end{tabular}
	%    \vspace{-0.5cm}
\end{table}

\begin{figure}[t]
	\centering
	\includegraphics[scale=.55]{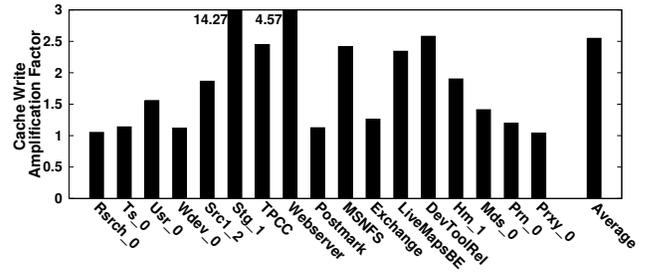}
	%\vspace{-0.7cm}
	\caption{CWAF for various workloads}
	\label{fig:cwaf}
	\vspace{-0.5cm}
\end{figure}

In \emph{Write-Back} (WB) cache policy which is commonly practiced in previous studies, each read miss requires writing a data page to the SSD while all write requests are directed to the SSD.
Therefore, the total number of writes in SSD will be higher than the number of write requests in the workload.
This will result in reduced lifetime of SSDs employed as a WB cache.
To evaluate the amplification of writes in previous studies, we introduce \emph{Cache Write Amplification Factor} (CWAF) parameter which is calculated based on Equation \ref{eq:cwaf}.
Fig. \ref{fig:cwaf} shows CWAF parameter for various workloads.
In \emph{Stg\_1} and \emph{Webserver} workloads, CWAF is greater than 4.5 which shows the importance of read misses on the SSDs lifetime.
By reducing the number of writes due to the read misses on SSDs, we can significantly improve the SSD endurance.

\begin{figure}[h]
	%\scriptsize
%	\vspace{-0.4cm}
	\begin{align}
	%\hspace{-.4cm}
	\label{eq:cwaf}
	CWAF &= \frac{Writes_{ssd}}{Writes_{workload}}
	\end{align}
	\vspace{-0.6cm}
\end{figure}

\begin{figure}
	\centering
	\includegraphics[scale=0.6]{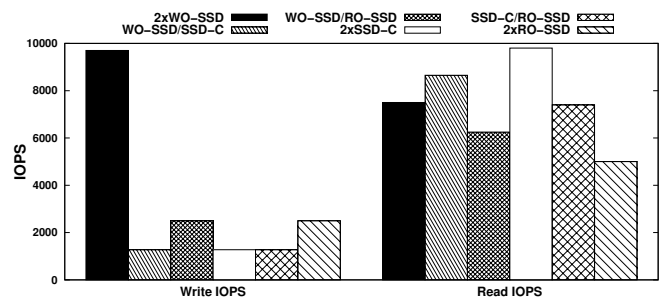}
	\caption{Performance of Heterogeneous RAID Architectures}
	\label{fig:raid}
	\vspace{-0.6cm}
\end{figure}

One of the reliability concerns of employing SSDs, specially in \emph{Redundant Array of Independent Disks} (RAID) configurations is correlated failures due to the either software or hardware defects \cite{Meza:2015:LSF:2745844.2745848}.
Since SSDs in the RAID configuration are identical and in mirrored RAID configurations they receive the same accesses, any software defect probably will trigger on both SSDs resulting in data loss.
Additionally, due to the same aging pattern and lifetime, both SSDs are expected to fail in a close time interval which also results in data loss.
To mitigate such problem and reduce the probability of double disk failures, employing heterogeneous SSDs with different internal algorithms and/or from different brands can be practiced.
Here, we investigate the effect of employing such technique on various MLC-TLC SSD combinations.
Fig. \ref{fig:raid} shows the normalized performance of various mirrored (RAID-1) configurations for heterogeneous SSDs compared to the performance of homogeneous mirrored SSDs.
As can be seen in this figure, the performance is limited by the slower SSD, specially in write requests which results in overall lower performance per cost.
For instance, replacing a SSD in a mirrored WO-SSD with a RO-SSD results in almost 5x performance degradation in write requests.
Write performance of two mirrored RO-SSDs is equal to the performance of mirrored WO-SSD and RO-SSD while the cost and power consumption of the latter architecture is higher.
In read requests, the performance degradation of employing heterogeneous architectures is lower compared to write requests since the performance gap of different SSDs is smaller in read requests.
This experiment shows that there is a need for heterogeneous SSD architectures with high performance per cost to simultaneously improve both performance and reliability.

\section{Proposed Architecture}
\label{sec:proposed}
An efficient I/O caching architecture should provide high performance, endurance, and reliability with reasonable cost overhead in order to be integrated in storage and high-performance servers.
Previous caching architectures have neglected to simultaneously consider such important parameters of I/O caching and focused on improving \emph{only} one of the parameters without investigating the corresponding impact on the other parameters.
The proposed architecture is motivated by the lack of a comprehensive caching architecture which 
is able to mitigate the shortcomings of previous studies discussed in Section \ref{sec:motivation}.
%is simultaneously optimized towards performance, endurance, and reliability.

For this purpose, we try to improve performance, power consumption, and lifetime of I/O cache by using a DRAM and high performance SSDs and reducing the number of committed writes to SSDs.
To address the reliability concern, TICA is designed such that it does not have any single point of failure and in addition, a failure in any of the caching devices will not result in data loss.
This is while the cost overhead is kept as small as possible compared to the traditional caching architectures.
TICA is also architected in such a way that the optimizations proposed in previous studies for increasing the cache hit ratio and prioritizing request types can be directly integrated with TICA in order to further improve performance and/or endurance.

\subsection{High-Level Architecture}
\begin{figure}
	\centering
	\includegraphics[scale=0.4]{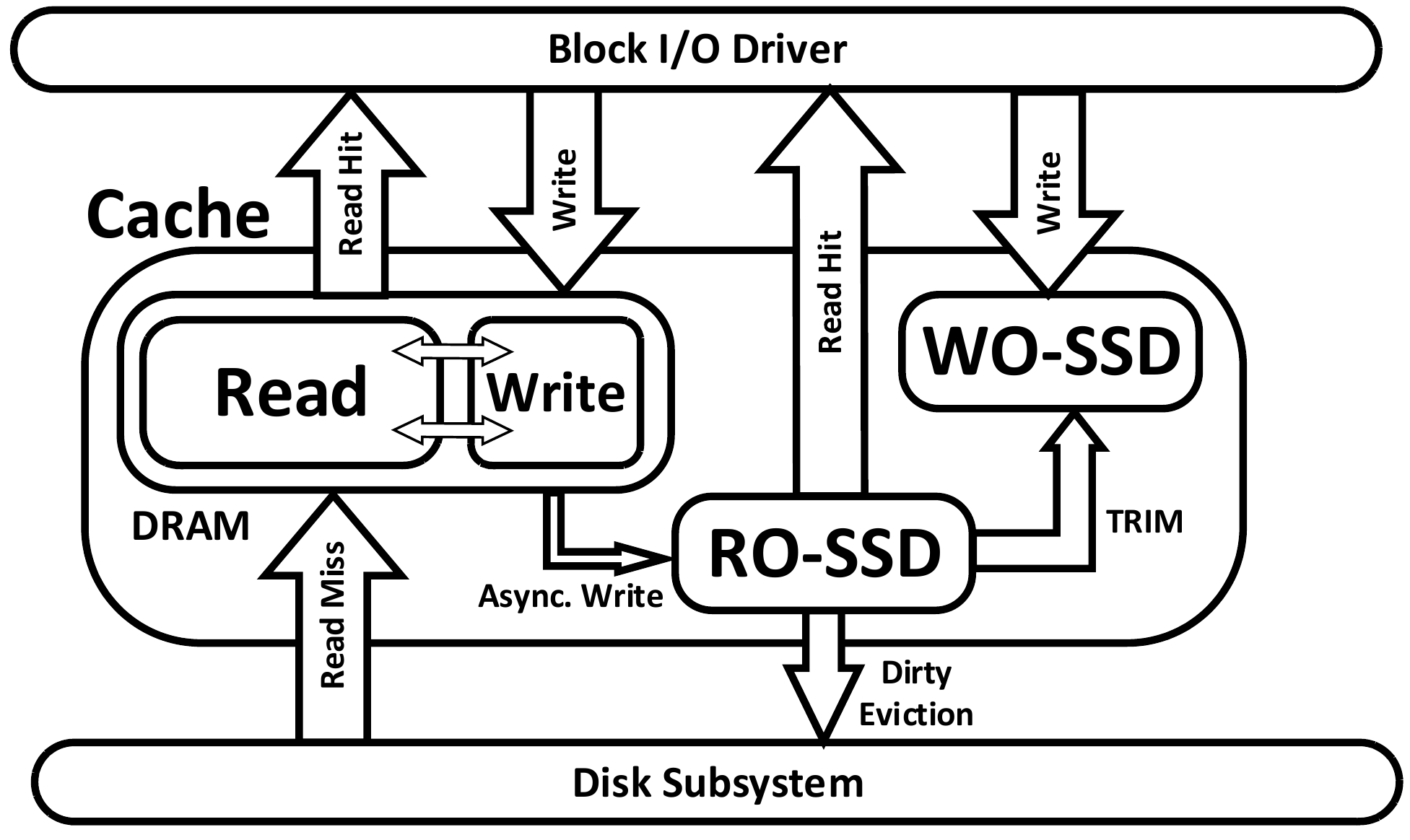}
	\caption{Proposed Architecture}
	\label{fig:arch}
	\vspace{-0.6cm}
\end{figure}

To design an efficient caching architecture, we leverage the traditional cache architecture and use three different storage devices for I/O caching.
%Since the policy for each device differs from other devices and the content of cache devices is not exactly mirrored, the proposed architecture has more flexibility than RAID 1 configuration.
A DRAM module alongside a RO-SSD and a WO-SSD form the three-levels of the proposed architecture.
In order to decrease the probability of data loss, a small battery-backup unit is added to DRAM which can sustain a cold system reboot.
Such heterogeneous architecture improves the reliability by reducing the probability of double disk failures due to the correlated failure between SSDs of the same model.
Fig. \ref{fig:arch} depicts the proposed architecture consists of three hardware modules.
The data migration inside the I/O cache or between the cache and the main storage device is done using \emph{Direct-Memory Access} (DMA) unit to reduce the CPU overhead.
Since a data page might exist in more than one caching device at any time, they are looked up based on the device priority which are prioritized as DRAM, RO-SSD, and then WO-SSD for read requests.
TICA works in write-back mode and as such, all write requests will be buffered.
If the old data page resides in any of the caching devices, it will be invalidated.
In addition to invalidation in mapping data structures, a TRIM\footnote{Informs disk about data blocks which are no longer in use by OS.} request is  sent to SSDs to improve its performance on write requests.

The proposed architecture also employs a DRAM in addition to SSDs in the caching layers where it is partitioned into read and asynchronous write cache sections.
The read cache partition is used for caching read miss requests.
The requested data page is moved to DRAM using DMA and afterwards the data page will be copied from DRAM cache to the destination memory address in the user space.
The user write requests arriving to the cache will be redirected to both WO-SSD and DRAM where they will be stored in the second partition of DRAM.
An asynchronous thread goes through the second partition and sends the data pages to the RO-SSD and removes them from DRAM.
The size of partitions is adjusted dynamically in the runtime based on the percentage of the write requests arrived to DRAM.

\begin{figure}
	\centering
	\includegraphics[scale=0.6]{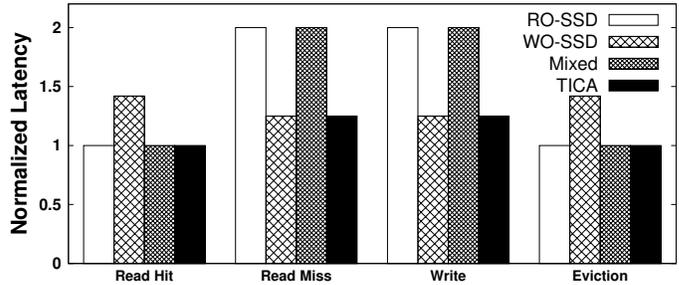}
	\caption{Average Response Time of Cache Operations Normalized to RO-SSD Read Latency}
	\label{fig:diffraid}
	\vspace{-0.6cm}
\end{figure}

To enhance the performance of the proposed architecture, RO-SSD and WO-SSD are configured in such a way that they reside in the critical path of responding to those requests that can be handled more efficiently.
This way TICA can have optimal performance on both read and write requests without having to use ultra high-performance SSDs which significantly reduces the total cost of I/O cache.
In order to show the difference between the proposed architecture and the traditional RAID 1 configurations, the normalized average response time under various cache operations is depicted in Fig. \ref{fig:diffraid}.
All configurations use two SSDs where in the first two configurations, SSDs in RAID 1 are the same and in the third configuration (mixed) and TICA, one RO-SSD and one WO-SSD are employed.
In order to have a fair comparison in Fig. \ref{fig:diffraid}, the DRAM module in the proposed architecture is ignored in this experiment.
As shown in Fig. \ref{fig:diffraid}, TICA has near optimal performance on every cache operation since the critical path of operations and the optimal operation for each SSD is considered.

\subsection{Detailed Algorithm}
Algorithm \ref{alg:caching} depicts the workflow of the proposed architecture in case of a request arrival.
If the request is to write a data page and the data page exists in the cache, it will be invalidated.
Lines \ref{alg:dramwcachefull} through \ref{alg:dramwcachefullend} check the DRAM write cache partition for free space.
If there is no space available, the size of the write cache partition will be extended.
The calculation for extending the write cache size considers a baseline cache size called $defwritecachesize$ and if the current write cache size is greater than this value, the write cache size will be extended by smaller values.
This technique prevents write cache partition from over extending which will reduce the number of read hits from DRAM.
In addition to DRAM, WO-SSD will be checked for free space and if there is no space left, a victim data page will be selected and discarded from both SSDs (lines \ref{alg:ssdwofull} through \ref{alg:ssdwofullend}).
The victim data page will be removed from RO-SSD since leaving a dirty data page in RO-SSD has a risk of data loss in case of failure of this SSD.
After allocating a page in both DRAM and WO-SSD, the write request will be issued.
The request for flushing from DRAM to RO-SSD will be issued after completion of the user request.

If an incoming request is for reading a data page, the caching devices will be searched based on their read performance (DRAM, RO-SSD, and WO-SSD, in order).
If the request is served from DRAM, $LRU_{DRAM}$ will be updated and if the request is hit in either of SSDs, the LRU queue for both SSDs will be updated.
If the request is missed in the cache while DRAM read cache is full and the DRAM write cache size is greater than $defwritecachesize$, the DRAM write cache size will be shrunk.
In order to shrink the write cache, it is required to wait for completion of one of the ongoing asynchronous writes to RO-SSD; this will make the current request being stalled.
On the other hand, evicting a data page from DRAM read cache imposes no I/O overhead.
Therefore, in the proposed architecture, a request is sent to the disk for reading the data page and a watcher thread waits for completion of one of the asynchronous writes to RO-SSD.
If one of the asynchronous writes is finished before disk I/O, its place will be allocated for the data page and if not, a data page will be evicted from DRAM read cache in order to make the required space available.
TICA allocates a few data pages from DRAM and SSDs for internal operations such as migrating data pages.
The default behavior of TICA is to discard the evicted data page from DRAM which we call TICA-EF.
There is an alternative approach which copies the evicted data page to WO-SSD which is called TICA-WED.
TICA-WED and the algorithm for selecting the TICA policy are detailed next.

\begin{algorithm}[t]
	\caption{Proposed Caching Algorithm}\label{alg:caching}
	\scriptsize
	\begin{algorithmic}[1]
		\Procedure{Access}{Request} \label{alg:access}
		\State capacityEstimator(Request)
		\If{Request.iswrite}
		\State IssueDiscards($Request.address$)
		\If {DRAMwritecache.isfull} \label{alg:dramwcachefull}
		\State \begin{varwidth}[t]{\linewidth}
			 $writecachesize \gets writecachesize +$\par
			 \hskip\algorithmicindent $2 ^ {-(writecachesize - defwritecachesize)}$
		\end{varwidth}
%		\State $writecachesize \mathrel{+}= 2 ^ {-(writecachesize - defwritecachesize)}$
		\State DiscardfromDRAM($writecachesize$)
		\State waitforFreeup
		\EndIf \label{alg:dramwcachefullend}
		\If {WOSSD.isfull} \label{alg:ssdwofull}
		\State FreeupWOSSD
		\State FreeupROSSD
		\EndIf \label{alg:ssdwofullend}
		\State Issue writes to $WOSSD$ and $DRAM$
		\State Wait for issued writes
		\State update $LRU_{DRAM}$ and $LRU_{WOSSD}$
		\State Issue async. write to $ROSSD$
		\Else
		\If {inDRAM($Request.address$)}
		\State ReadfromDRAM($Request.address$)
		\State Update $LRU_{DRAM}$
		\ElsIf {InROSSD($Request.address$)}
		\State ReadfromROSSD($Request.address$)
		\State Update $LRU_{ROSSD}$
		\State Update $LRU_{WOSSD}$
		\ElsIf {InWOSSD($Request.address$)}
		\State ReadfromWOSSD($Request.address$)
		\State Update $WOSSDLRU$
		\Else 
		\If{DRAMReadcache.isfull}
		\State \begin{varwidth}[t]{\linewidth}
			writecachesize $\gets$ max($defwritecachesize$,\par
			\hskip\algorithmicindent $writecachesize  \penalty 0 -  2 ^ {writecachesize - defwritecachesize}$)
			%		writecachesize \gets max ( $defwritecachesize$,\par
			%							\hskip\algorithmicindent $writecachesize  \penalty 0 -  2 ^ {writecachesize - defwritecachesize}$)
		\end{varwidth}
		\State DiscardfromDRAM($writecachesize$)
		\If {TICA is in \textbf{WED} mode}
		\State Copy evicted page to \textbf{WOSSD}
		\EndIf
		\State Issue page fault for $Request.address$
		\EndIf
		\EndIf
		\EndIf
		\EndProcedure \label{alg:lookupend}
	\end{algorithmic}
\end{algorithm}

\subsection{TICA-EF vs. TICA-WED}
\label{sec:tica-wed}
\begin{figure}
	\centering
	\includegraphics[scale=.57]{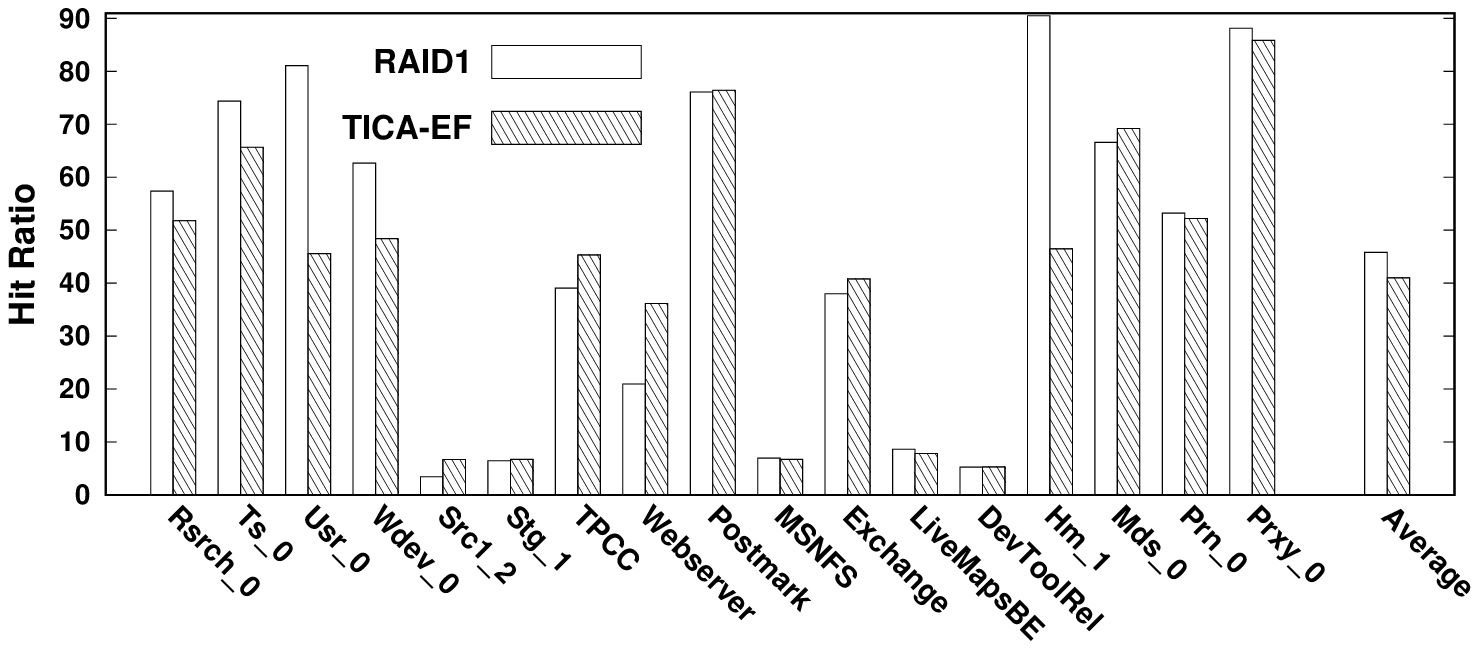}
	%\vspace{-0.7cm}
	\caption{TICA-EF Hit Ratio}
	\label{fig:efwed-hit}
	\vspace{-0.6cm}
\end{figure}

As mentioned earlier, the endurance of the SSD caching architectures is penalized by the read misses.
TICA-EF eliminates the writes in the SSDs due to the read misses and therefore, is called \emph{Endurance-Friendly}.
Such approach, however, imposes performance cost since the data pages are evicted early from the cache and cache hit ratio is decreased.
Fig. \ref{fig:efwed-hit} shows the evaluation of the TICA-EF in terms of the cache hit ratio compared to the baseline RAID-1 configuration.
TICA-EF fails to provide high performance in several workloads such as \emph{Usr\_0}, \emph{Hm\_1}, and \emph{Wdev\_0}.
Our investigation reveals that this is due to the large working set size of the read-intensive data pages.
Such data pages can be only buffered in DRAM and since DRAM has a small size, data pages are evicted before re-referencing.
Therefore, TICA-EF needs to access HDD more often to bring back evicted data pages to DRAM.

\begin{figure}
	\centering
	\includegraphics[scale=.57]{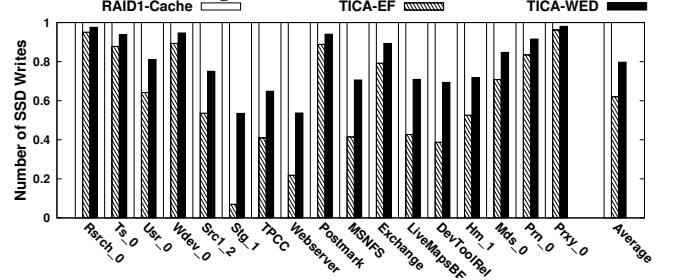}
	%\vspace{-0.7cm}
	\caption{Total Number of Writes Committed to SSDs}
	\label{fig:efwed-writes}
	\vspace{-0.6cm}
\end{figure}

Copying the evicted data pages from DRAM to SSD can improve performance at the cost of reducing endurance.
To show the effect of such technique, we propose TICA-WED which copies the data pages on eviction from DRAM to WO-SSD.
As mentioned in the motivation section (Section \ref{sec:motivation}), this will decrease the endurance of SSDs.
Fig. \ref{fig:efwed-writes} shows the number of writes committed to SSDs in TICA-WED compared to TICA-EF.
In read-intensive workloads with small working set size, TICA-EF has close endurance efficiency to TICA-WED.
In other workloads, however, TICA-WED has higher endurance efficiency.
We can conclude here that both TICA-EF and TICA-WED policies can provide a suitable policy for a specific workload type and a general approach is required to select one of these two policies based on the workload characteristics.

\subsection{Adaptive TICA}
\label{sec:adaptive}
To select an effective policy for TICA, we have analyzed the performance of TICA-EF.
The performance behavior of TICA-EF in \emph{Wdev\_0}, \emph{Usr\_0}, \emph{Ts\_0} and \emph{Rsrch\_0} workloads reveals that there are two reasons for low performance of TICA-EF: 1) DRAM size is less than the working set size, and 2) cold data pages are trapped in SSDs.
To mitigate the performance degradation of TICA-EF, we propose \emph{TICA-Adaptive} (TICA-A) which switches the policy from TICA-EF to TICA-WED when one of the two above conditions is detected.
In this section, the algorithms for identifying the mentioned two conditions are detailed.

\subsubsection{DRAM Low Capacity Identifier}
\label{sec:dramlow}
Due to the small size of DRAM in TICA, the thrashing problem \cite{Seshadri:2012:EFU:2370816.2370868} is likely to happen if the working set size of the workload is larger than DRAM size.
To identify such condition, we keep a queue of evicted data pages from DRAM, called \emph{Evicted Queue} (EQ).
Evicted data pages from DRAM enter EQ if they are copied to the DRAM due to a read miss.
The hit ratio \emph{with} and \emph{without} considering the EQ is calculated periodically and if their difference is greater than a predefined threshold ($T_{min}$), TICA will switch to the TICA-WED policy.
Employing the threshold for minimum difference between hit ratios prevents constantly switching between the two policies.

Since TICA-EF lowers the \emph{Total Cost of Ownership} (TCO) by extending the SSDs lifetime, we prefer it over TICA-WED.
Therefore, if TICA-EF has high hit ratio, regardless of the hit ratio of EQ, we switch to TICA-EF.
The threshold ($T_{max}$), however, should be set conservatively to ensure negligible performance degradation.
Modifying the thresholds enables us to prefer one of the two policies based on the I/O demand of the storage system and/or the overall system status.
For instance, when most SSDs in the array are old, we would prefer TICA-EF to prolong their lifetime and reduce the probability of data loss.
Algorithm \ref{alg:capacity} shows the flow of identifying thrashing in DRAM.
Switching between the policies is conducted once the number of incoming requests to the cache becomes twice the size of DRAM memory.
For each request, the counter for hits in DRAM and EQ are updated in Lines \ref{eq:hitstart} through \ref{eq:hitend}.
In Lines \ref{eq:checkstart} to \ref{eq:checkend}, the hit ratios are checked and TICA policy is changed if required.

\begin{algorithm}[t]
	\caption{DRAM Low Capacity Identifier}\label{alg:capacity}
	\scriptsize
	\begin{algorithmic}[1]
		\State $windowSize \gets 2 * DRAM_{size}$
		\State $requestCounter, EQHit, DRAMReadHit \gets 0$
		\Procedure{capacityEstimator}{request}
		\State $requestCounter \gets requestCounter + 1$
		\If{request.isRead} \label{eq:hitstart}
		\If{Hit in DRAM}
		\State $DRAMReadHit \gets DRAMReadHit + 1$
		\ElsIf{Hit in EQ}
		\State $EQHit \gets EQHit + 1$
		\EndIf
		\EndIf \label{eq:hitend}
		\If{$requestCounter == windowSize$}
		\If{$(EQHit+DRAMReadHit)>T_{max}$} \label{eq:checkstart}
		\State Switch to \textbf{TICA-WED}
		\ElsIf{$EQHit>T_{min}$}
		\State Switch to \textbf{TICA-WED}
		\Else
		\State Switch to \textbf{TICA-EF}
		\EndIf
		\State $requestCounter, EQHit, DRAMReadHit \gets 0$ \label{eq:checkend}
		\EndIf 
		\EndProcedure
	\end{algorithmic}
\end{algorithm}

\subsubsection{Preventing Cold Data Trapped in SSDs}
\label{sec:coldssd}
\begin{figure}
	\centering
	\includegraphics[scale=.4]{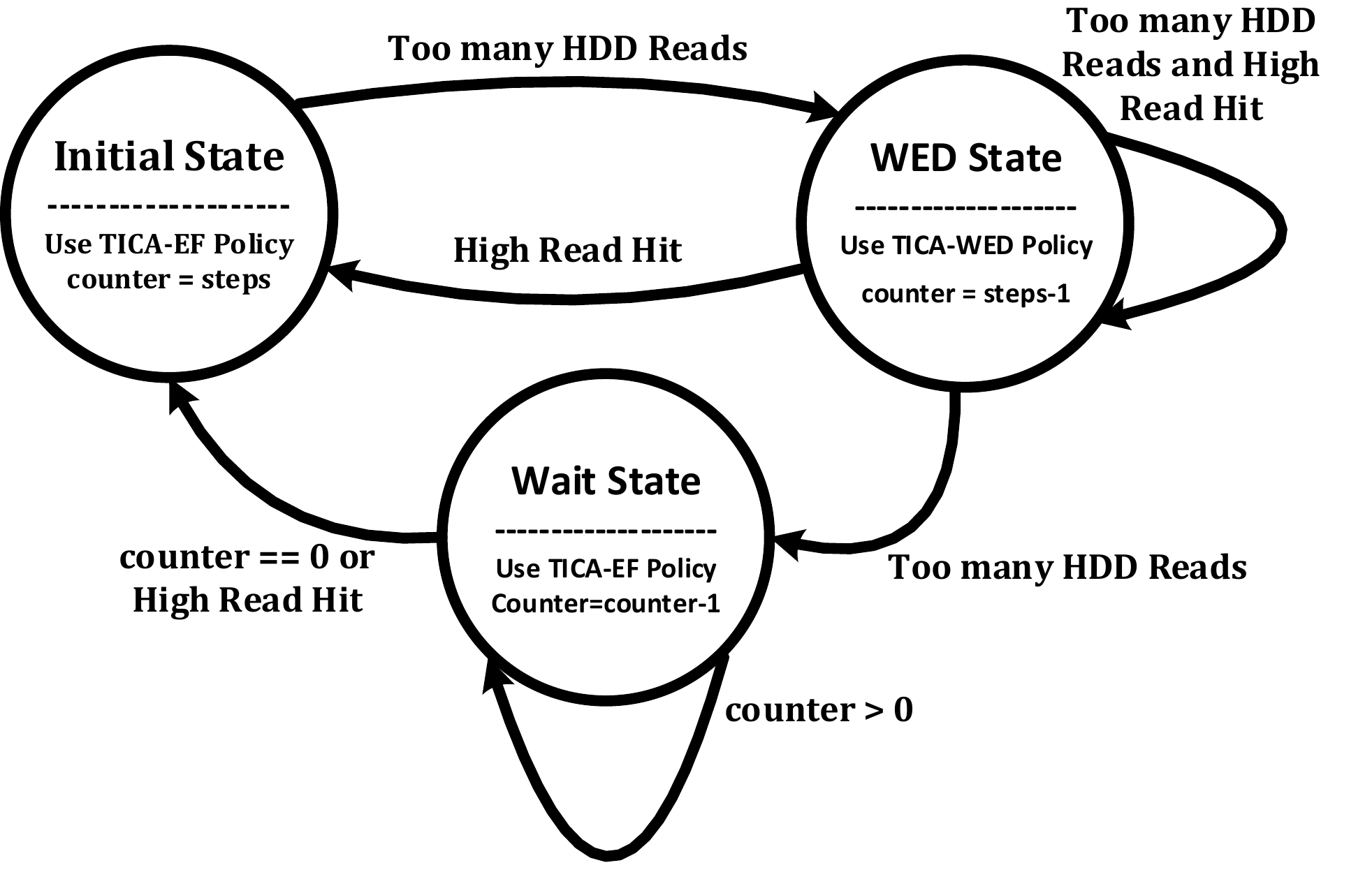}
	%\vspace{-0.7cm}
	\caption{State Machine for Preventing Cold Data Trapped in SSD}
	\label{fig:smbi}
	%\vspace{-0.6cm}
\end{figure}

In TICA-EF, only write accesses are redirected to SSDs and all read accesses are supplied by either DRAM or HDD.
Therefore, in read-intensive workloads, SSDs become idle and previously hot data pages which are now cold reside in SSDs without any means to evict such data pages.
To prevent such problem, we propose a \emph{State Machine Based Insertion} (SMBI) to \emph{conservatively} switch from TICA-EF to TICA-WED in order to replace the cold data pages in SSDs.
The simplified model of SMBI is shown in Fig. \ref{fig:smbi}.
We identify two conditions 1) too many HDD reads and 2) high hit ratio. When both conditions are met in the workload, TICA switches to TICA-WED until one of the conditions is no longer valid.
\emph{Too many HDD reads} shows that the read working set size is larger than DRAM size.
In such condition, we allow evicted data pages from DRAM to enter WO-SSD to increase its hit ratio and reduce the number of HDD reads.
We cannot rely solely on the number of HDD reads for switching to WED since in workloads with low locality, the number of HDD reads is also high and copying the evicted data pages from DRAM to WO-SSD will only impose endurance cost without any performance improvement.
Therefore, SMBI stays in the WED state as long as both number of HDD reads and hit ratio are high.
Having high hit ratio and low number of HDD reads shows that the working set size is smaller than 
DRAM size and SMBI switches back to the EF policy.

If the hit ratio is decreased while the number of HDD reads is still high, SMBI enters a waiting state which prevents re-entering WED mode in the next few windows.
This state prevents constantly switching between the two policies.
Algorithm \ref{alg:smbi} shows the detailed flow of SMBI.
Line \ref{alg:thdd} switches the policy to TICA-WED if the number of HDD read requests in the current window is greater than the $T_{hdd}$ threshold.
In lines \ref{alg:wedstart} through \ref{alg:wedend}, SMBI checks both conditions and if one of them is no longer valid, it switches back to TICA-EF policy.

\begin{algorithm}[t]
	\caption{State Machine Based Insertion} \label{alg:smbi}
	\scriptsize
	\begin{algorithmic}[1]
		\State $counter \gets steps$
		\State $ currentState \gets initialState, nextState \gets initialState$
		\State $diskRead, readHit, requestCounter \gets 0$
		\Procedure{SMBI}{request}
		\State $requestCounter \gets requestCounter + 1$
		\If{request.isRead}
		\If{Hit in cache}
		\State $readHit \gets readHit + 1$
		\Else
		\State $diskRead \gets diskRead + 1$
		\EndIf
		\EndIf
		\If{$requestCounter == sampleSize$}
		\If{$currentState == initialState$}
		\If{$diskRead>T_{hdd}$} \label{alg:thdd}
		\State Switch to \textbf{TICA-WED} policy
		\State $counter \gets steps - 1$
		\State $currentState \gets WEDState$
		\EndIf
		\ElsIf {$currentState == WEDState$}
		\If {$diskRead>T_{hdd}$} \label{alg:wedstart}
		\If {$readHit>T_{read}$}
		\State Switch to \textbf{TICA-WED} policy
		\State $currentState \gets WEDState$
		\Else
		\State Switch to \textbf{TICA-EF} policy
		\State $currentState \gets waitState$
		\EndIf
		\ElsIf {$readHit>T_{read}$}
		\State Switch to \textbf{TICA-EF} policy
		\State $counter \gets steps$
		\State $currentState \gets initialState$ \label{alg:wedend}
		\EndIf
		\ElsIf {$currentState == waitState$}
		\If {$(counter == 0 \textbf{ or } readHit)>T_{read}$}
		\State Switch to \textbf{TICA-EF} policy
		\State $counter \gets steps$
		\State $currentState \gets initialState$
		\Else
		\State $counter \gets counter - 1$
		\EndIf
		\EndIf
		\EndIf
		\EndProcedure
	\end{algorithmic}
\end{algorithm}

\begin{figure}[t]
	\centering
	\includegraphics[scale=.4]{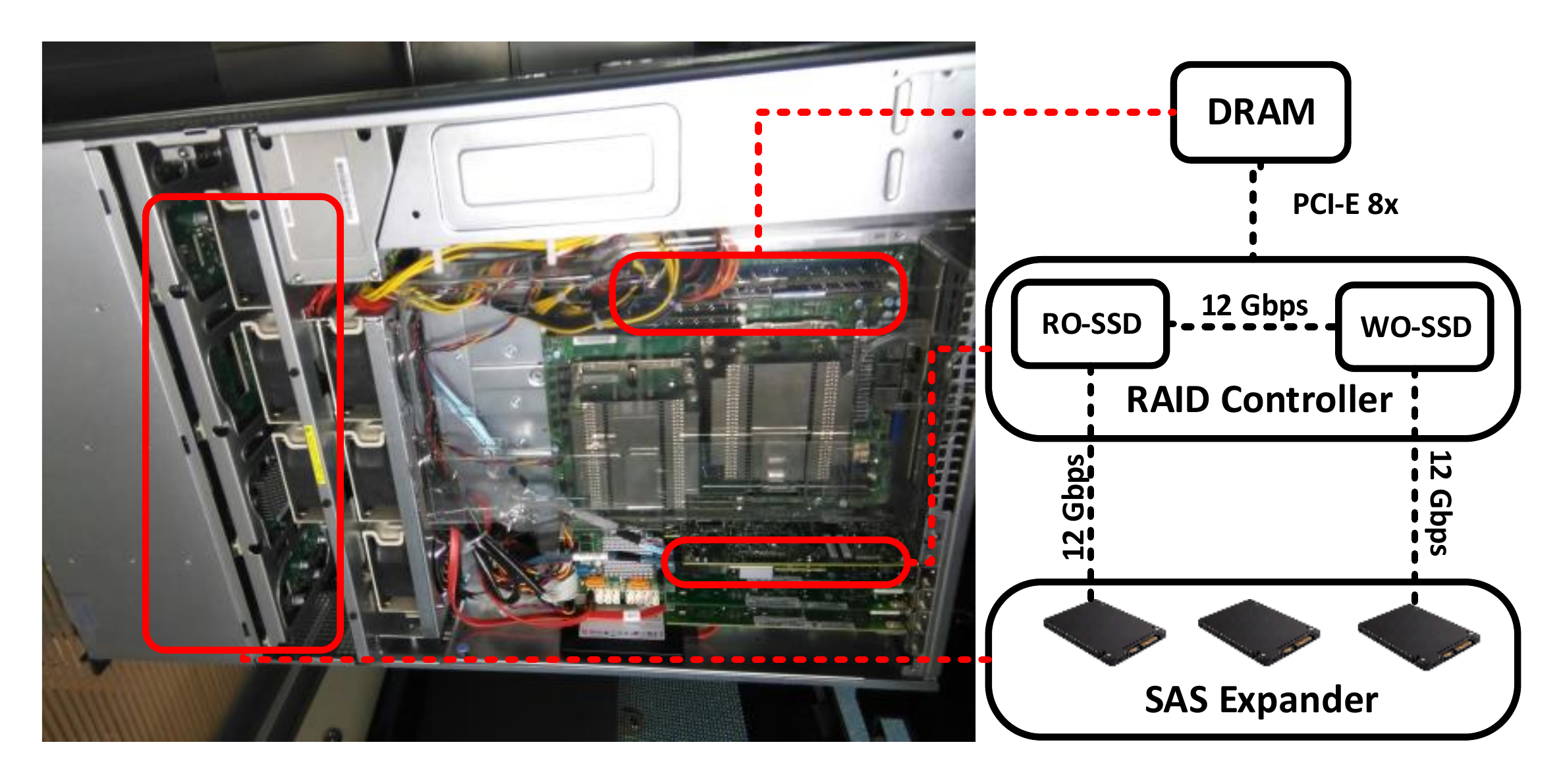}
	\vspace{-0.3cm}
	\caption{Hardware Architecture of Experimental Setup}
	\label{fig:hardware}
	%\vspace{-0.6cm}
\end{figure}

\begin{figure*}
%	\noindent\colorbox{amber}{
%		\parbox{\textwidth}
%		{
	\centering
	\includegraphics[scale=1.23]{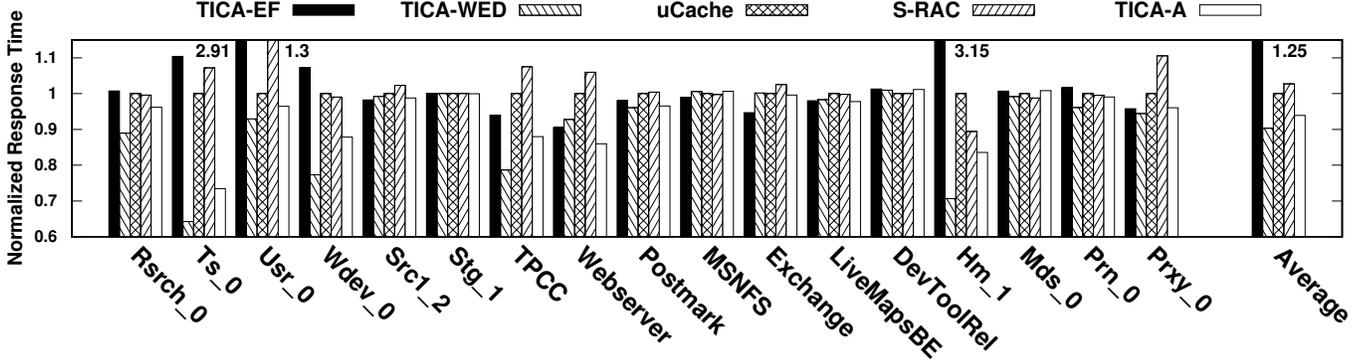}
	%\vspace{-0.7cm}
	\caption{Normalized Response Time: TICA vs. Conventional Caching Architectures}
	\label{fig:performance}
	%\vspace{-0.6cm}
%}
%}
\vspace{-.3cm}
\end{figure*}

%\subsection{Power Consumption Model}
%\vspace{-0.1cm}

%\subsection{Reliability Model}
%\vspace{-0.1cm}
%To evaluate reliability of TICA and compare it with baseline architectures, a reliability metric called \emph{Bits Lost Per Year} (BLPY) is proposed.
%BLPY is the expected value of bits lost per gigabyte in a year for a given storage device based on its \emph{Mean Time to Failure} (MTTF).
%Such parameter is selected for evaluating reliability over Markov processes since TICA employs an additional storage device (DRAM) compared to baseline architectures and does not follow simple RAID configurations.
%BLPY is computed similarly to MTTDL \cite{Greenan:2010:MTM:1863122.1863127} which is shown to be an accurate parameter for comparing reliability of storage systems.
%
%\begin{figure}[h]
%	\scriptsize
%	\vspace{-0.4cm}
%	\begin{align}
%	\hspace{-.4cm}
%	\label{eq:reliability}
%	BLPY_{TICA} = &BLPY_{DRAM,WO-SSD} + BLPY_{WO-SSD,RO-SSD} \\ \nonumber
%	& - BPLY_{DRAM,WO-SSD,RO-SSD}
%	\end{align}
%	\vspace{-0.4cm}
%\end{figure}
%
%TICA will have data loss in two scenarios, a) DRAM and WO-SSD failure, and b) WO-SSD and RO-SSD failure.
%In both cases, data pages equal to the size of WO-SSD will be lost (considering RO-SSD with the same size as WO-SSD).
%Two scenarios have different BLPY since DRAM does not have the same MTTF as SSDs.
%Reliability of TICA will be computed based on Equation \ref{eq:reliability}.
%Since the extreme rare case of failure of all three storage devices is included in both scenarios
%and therefore is summed twice, it is decremented once in the formula.

\vspace{-0.2cm}
\section{Experimental Results}
\label{sec:results}
In this section, the performance, power consumption, endurance, and reliability of the proposed architecture is evaluated.
We compare TICA with a state-of-the-art multi-level caching architecture (uCache \cite{6831945}) {and a state-of-the-art SSD caching architecture (S-RAC \cite{Ni:2016:SSF:2928275.2928284}).}
In order to have a fair comparison, uCache is modified and all single points of failure are removed to improve its reliability.
Support for RAID1 SSDs is also added to uCache.
{Since S-RAC is a single-level cache architecture, a first level DRAM cache 
is added so that all three examined architectures benefit from both DRAM and 
SSD.}
To show the effect of different TICA policies, in addition to TICA-A, both TICA-EF and TICA-WED are also evaluated.
The detailed characteristics of the workloads are reported in Table \ref{tbl:workloads}.

\vspace{-.3cm}
\subsection{Experimental Setup}
\label{sec:expsetup}
To conduct the experiments, we employed a rackmount server equipped with Intel Xeon, 128GB memory, and a SSD for the operating system to reduce the effect of the operating system and the other running applications on the obtained results.
Fig. \ref{fig:hardware} shows the actual server running experiments and the interfaces between I/O cache layers.
\emph{SAS expander} is capable of supporting both SATA and SAS disk drivers.
\emph{RAID controller} is configured in \emph{Just a Bunch Of Disks} (JBOD) mode where disks are directly provided to the OS without any processing by the controller.
Employing \emph{SAS expander} and \emph{RAID controller} enables us to run experiments on various SATA/SAS SSDs without need for disk replacement or server reboot.

WO- and RO-SSDs are selected from enterprise-grade SSDs employed in the datacenters.
We warm up SSDs before each experiment by issuing requests until the SSD reaches a stable latency.
The traces are replayed on the devices using our in-house trace player which is validated by blktrace \cite{blktrace} tool.
The requests sent to the disk by our trace player are compared to the original trace file to ensure it has the expected behavior.
The characteristics of DRAM and SSDs employed in the experimental results is reported in Table \ref{tab:cachingdevices}.
In the experiments, size of SSDs and DRAM is set to 10\% and 1\% of the working set size, respectively.
The value of $T_{min}$, $T_{max}$, $T_{hdd}$, and $T_{read}$ are set to 0.15, 0.25, 0.2, and 0.2, respectively.

\begin{table}[t]
	\caption{Workload Characteristics}
	\scriptsize
	\label{tbl:workloads}
	\centering
	\begin{tabular}{|l|c|c|c|}
		\hline
		\textbf{Workload}   & \textbf{\shortstack{Total \\ Requests Size}} & \textbf{Read Requests} & \textbf{Writes Requests}\\ \hline
		TPCC       & 43.932 GB           & 1,352,983 (70\%)       & 581,112 (30\%) \\ \hline
		Webserver  & 7.607 GB            & 418,951 (61\%)      & 270,569 (39\%)   \\ \hline
		DevToolRel & 3.133 GB            & 108,507 (68\%)        & 52,032 (32\%)   \\ \hline
		LiveMapsBE & 15.646 GB           & 294,493 (71\%)       & 115,862 (28\%)  \\ \hline
		MSNFS      & 10.251 GB           & 644,573 (65\%)       & 349,485 (35\%)     \\ \hline
		Exchange   & 9.795 GB            & 158,011 (24\%)        & 502,716 (76\%)  \\ \hline
		Postmark   & 19.437 GB           & 1,272,148 (29\%)       & 3,172,014 (71\%) \\ \hline
		Stg\_1     & 91.815 GB           & 1,400,409 (64\%)       & 796,452 (36\%)    \\ \hline
		Rsrch\_0   & 13.11 GB            & 133,625 (9\%)       & 1,300,030 (91\%)     \\ \hline
		Src1\_2    & 1.65 TB          & 21,112,615 (57\%)     & 16,302,998 (43\%)   \\ \hline
		Wdev\_0    & 10.628 GB           & 229,529 (20\%)       & 913,732 (80\%)    \\ \hline
		Ts\_0      & 16.612 GB           & 316,692 (18\%)       & 1,485,042 (82\%)   \\ \hline
		Usr\_0     & 51.945 GB           & 904,483 (40\%)       & 1,333,406 (60\%)   \\ \hline
		Hm\_1    & 9.45 GB          & 580,896 (94\%)     & 28,415 (6\%)   \\ \hline
		Mds\_0    & 11.4 GB           & 143,973 (31\%)       & 1,067,061 (69\%)    \\ \hline
		Prn\_0      & 63.44 GB           & 602,480 (22\%)       & 4,983,406 (78\%)   \\ \hline
		Prxy\_0     & 61.03 GB           & 383,524 (5\%)       & 12,135,444 (95\%)   \\ \hline
	\end{tabular}
\end{table}

\vspace{-.3cm}
\subsection{Performance}
\label{sec:performance}
Fig. \ref{fig:performance} shows the normalized response time of TICA compared 
to uCache {and S-RAC, all normalized to uCache}.
TICA-WED which is optimized toward higher performance, reduces the response 
time by {12\%} on average compared to uCache {and S-RAC}.
The highest performance improvement {of TICA} belongs to \emph{Ts\_0} 
workload with {45\%} reduction in response time {(compared to S-RAC)}.
Although TICA-WED and TICA-EF differ in read miss policy and \emph{Ts\_0} is a 
write-dominant workload (80\% write requests), TICA-WED still performs better 
than TICA-EF with 42\% less response time.
This shows the significant impact of writing read misses on the SSDs and therefore, forcing the dirty data pages to be evicted from cache.
TICA-WED also improves performance in read-dominant workloads such as TPCC by copying the evicted data pages from DRAM to WO-SSD.

TICA-EF, optimized toward better endurance, outperforms TICA-WED in few workloads such as \emph{Webserver} and \emph{Exchange}.
Our investigation reveals that this is due to a) the limited space of SSDs and b) forcing the eviction of dirty data pages from SSD which is conducted aggressively in TICA-WED.
In \emph{Webserver} workload, TICA-A also identifies such problem and manages copying evicted data pages from DRAM to WO-SSD.
Therefore, it has better performance-efficiency in \emph{Webserver} workload compared to both TICA-EF and TICA-WED policies.
By managing the evicted data pages from DRAM, TICA-A improves performance 
compared to {previous studies by up to 45\% and 8\% on average.}
We can conclude here that TICA-A is performance-efficient in both read- and write-intensive workloads by managing the evicted data pages from DRAM.

%Fig. \ref{fig:performance} shows the average response time of various cache architectures normalized to the response time of the SSD-C RAID1.
%As shown in Fig. \ref{fig:performance}, WO-SSD demonstrates higher performance in all benchmarks compared to RO-SSD (by an average of 50\%).
%This is due to the relatively close performance of WO-SSD comapred to RO-SSD in read requests.
%Since SSDs have a high performance on read requests, even write-optimized SSDs have a high read performance.
%Therefore, the gain achieved by optimizing SSDs towards write requests is much more than the gain of optimizing them towards read requests.
%On the other hand, write-optimized SSDs are more expensive and increase the total cost of the cache.
%In the proposed technique, both RO-SSD and WO-SSD are employed which reduces the cost compared to the WO-SSD RAID 1 and improves the performance compared to RO-SSD and WO-SSD RAID1 architectures.
%TICA reduces AMAT by 65\% and 31\% on average compared to RO-SSD and WO-SSD, respectively.
%Using a DRAM read cache and redirecting requests to the optimized SSD techniques result in a significant performance improvement.
%Since TICA issues asynchronous write requests to RO-SSD, this SSD can be replaced with SSD-C to further reduce the cost overhead (by 4.5\%) while reducing the performance gain from 49\% to 44\%.
%This has been reported in Fig. \ref{fig:performance} as TICA-SSD-C.

\begin{figure*}
%	\noindent\colorbox{amber}{
%		\parbox{\textwidth}
%		{
	\centering
	\includegraphics[scale=1.23]{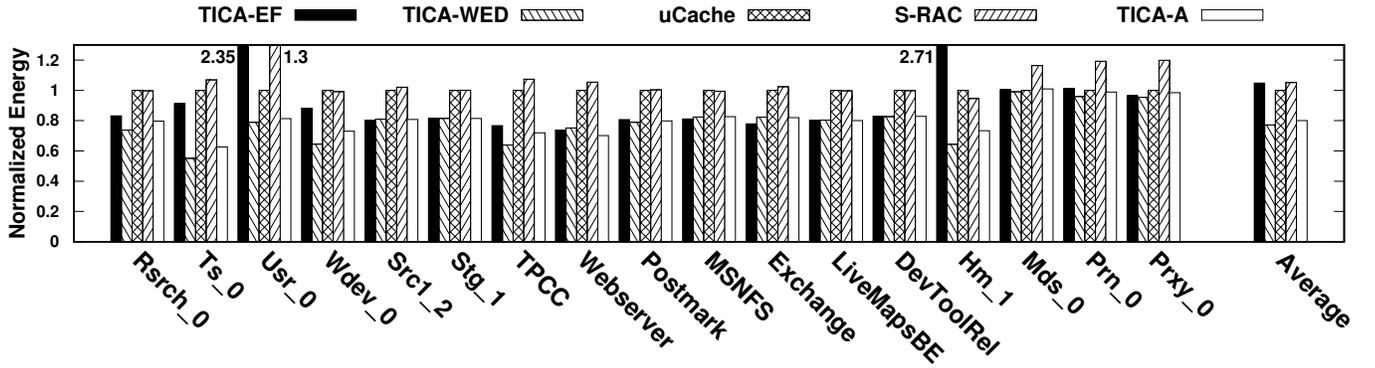}
	%\vspace{-0.7cm}
	\caption{Normalized Power Consumption: TICA vs. Conventional Caching Architectures}
	\label{fig:power}
	
%}
%}
\vspace{-0.45cm}
\end{figure*}

\vspace{-.3cm}
\subsection{Power Consumption}
To evaluate the power consumption of TICA, we estimate the total consumed energy for workloads.
In addition to the read and write requests, idle power consumption of the devices is also considered in the energy consumption to further increase its accuracy.
The read and write operations for background tasks such as copying the dirty data pages from DRAM to RO-SSD and flushing such data pages from RO-SSD to disk are also included in the energy consumption formula.
Equation \ref{eq:power} shows the formula for estimating the total energy consumption.
All parameters are detailed in Table \ref{tbl:paramdesc}.

\begin{figure}[h]
	\scriptsize
	\vspace{-0.4cm}
	\begin{flalign}
	\hspace{-.5cm}
	\label{eq:power}
	Energy&=\nonumber\\
	&\sum\limits_{}^{Read_{wo}}(RLat_{wo}*RP_{wo})+\sum\limits_{}^{Write_{wo}}(WLat_{wo}*WP_{wo})+\nonumber\\&(Idle_{wo}*IP_{wo})+
	\sum\limits_{}^{Read_{ro}}(RLat_{ro}*RP_{ro})+\nonumber\\&\sum\limits_{}^{Write_{ro}}(WLat_{ro}*WP_{ro})+(Idle_{ro}*IP_{ro})+\nonumber\\
	&\sum\limits_{}^{(Read_{D}+Write_{D})}(Lat_{D}*P_{D})+(Idle_{D}*IP_{ro})
	\end{flalign}
	\vspace{-0.4cm}
\end{figure}

\begin{table}[t]
	\caption{Parameters Description}
	\label{tbl:paramdesc}
	\vspace{-.3cm}
	\scriptsize
	\centering
	\hspace{-.5cm}
	\begin{tabular}{|c|c|}
		\hline
		\textbf {Parameter}  & \textbf{Description} \\ \hline   
		$Read_{wo}$ & WO-SSD Total Read Requests\\ \hline
		$Write_{wo}$ & WO-SSD Total Write Requests\\ \hline
		$Read_{ro}$ & RO-SSD Total Read Requests\\ \hline
		$Write_{ro}$ & RO-SSD Total Write Requests \\ \hline
		$Read_{D}$ & DRAM Total Read Requests \\ \hline
		$Write_{D}$ & DRAM Total Write Requests \\ \hline
		$RLat_{wo}$ & WO-SSD Read Latency \\ \hline
		$WLat_{wo}$ & WO-SSD Write Latency \\ \hline
		$RLat_{ro}$ & RO-SSD Read Latency \\ \hline
		$WLat_{ro}$ & RO-SSD Write Latency  \\ \hline
		$Lat_{wo}$ & DRAM Latency \\ \hline
		$RP_{wo}$ & WO-SSD Read Power \\ \hline
		$WP_{wo}$ & WO-SSD Write Power \\ \hline
		$RP_{ro}$ & RO-SSD Read Power  \\ \hline
		$WP_{ro}$ & RO-SSD Write Power \\ \hline
		$P_{D}$ & DRAM Power \\ \hline
		$IP_{wo}$ & WO-SSD Idle Power \\ \hline
		$IP_{ro}$ & RO-SSD Idle Power \\ \hline
		$IP_{D}$ & DRAM Idle Power \\ \hline
		$Idle_{wo}$ & Total WO-SSD Idle Time \\ \hline
		$Idle_{ro}$ & Total RO-SSD Idle Time \\ \hline
		$Idle_{D}$ & Total DRAM Idle Time \\ \hline
		$R_{Device}$ & Device Reliability \\ \hline
		$U_{Device}$ & Device Unreliability \\ \hline
		$MTTF_{Device}$ & Device Mean Time To Failure \\ \hline
	\end{tabular}
	\vspace{-.3cm}
\end{table}

\begin{figure*}[t]
%\noindent\colorbox{amber}{
%\parbox{\textwidth}
%{
	\centering
	\includegraphics[scale=1.2]{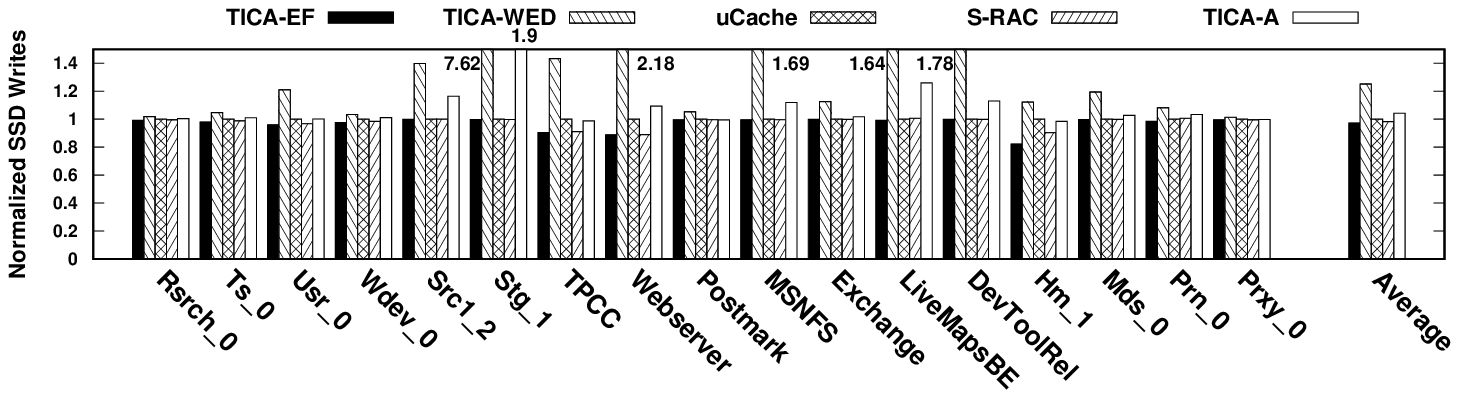}
	%\vspace{-0.3cm}
	\caption{Normalized Number of Writes Committed to SSDs}
	\label{fig:writes}
	%\vspace{-0.6cm}
%}
%}
\vspace{-.3cm}
\end{figure*}

TICA improves the power consumption by a) employing power-efficient SSDs while maintaining the performance and b) reducing the number of accesses to the SSDs.
Previous studies employ two identical SSDs in a mirrored RAID configuration to provide high reliability while as discussed in Section \ref{sec:motivation}, heterogeneous SSDs are not performance-efficient in traditional mirrored RAID configurations.
As such, state-of-the-art architectures such as uCache {and S-RAC} need to 
employ two WO-SSDs, which have high power consumption.
TICA on the other hand, employs a WO-SSD and a RO-SSD in its architectures which results in lower power consumption compared to using two WO-SSDs.
Additionally, by reducing the response time of the requests, SSDs more often enter idle state and therefore, the total power consumption is decreased.
Fig. \ref{fig:power} shows the normalized consumed energy of TICA compared to 
uCache {and S-RAC, normalized to uCache}.
In all workloads, TICA policies improve power consumption which shows the effectiveness of replacing a WO-SSD with a RO-SSD.
The only {exceptions are \emph{Usr\_0} and \emph{Hm\_1} workloads where 
TICA-EF has 2.35x and 2.71x higher power consumption compared to uCache, 
respectively.}
This is due to the high response time of the requests in this workload, which 
prevents SSDs from entering the idle state.
TICA-A improves power consumption by an average of {28\%} and the maximum 
improvement in the power consumption {(70\%)} belongs to \emph{Ts\_0} workload 
{in comparison with S-RAC}.

%RO-SSD has lower power consumption in both read and write requests, however, due to having longer delay, it consumes more energy
%in write requests compared to WO-SSD which will increase power consumption of TICA.
%Responding requests from DRAM, however, reduces the total power consumption of TICA.
%Normalized EPR for various cache architectures is depicted in Fig. \ref{fig:power}.
%In write-dominant workloads, TICA slightly consumes more energy compared to WO-SSD RAID1.
%This is due to employing RO-SSD in TICA which is not as energy efficient as WO-SSD.
%The average power consumption of TICA, however, is 11\% lower than WO-SSD RAID1.
%Highest power consumption improvement belongs to the \emph{stg} workload due to having low hit ratio.
%Baseline algorithms copy data pages from disk to SSDs in read misses which requires a write operation on SSDs.
%TICA, however, copies such data pages to DRAM which consumes significantly less energy.

\vspace{-.3cm}
\subsection{Endurance}
TICA-WED redirects all evicted data pages from DRAM to WO-SSD and therefore, significantly increases the number of writes in SSDs.
TICA-EF on the other hand, does not copy such data pages to SSDs to preserve their lifetime.
Fig. \ref{fig:writes} shows the normalized number of writes in the SSDs 
compared to uCache {and S-RAC, normalized to uCache}.
{uCache, S-RAC, and TICA-EF} have almost the same number of writes in SSDs 
since 
{they} limit writes in the SSDs.
TICA-EF improves SSDs lifetime by an average of {1.3\%} compared to uCache 
{and S-RAC}.

TICA-WED places all evicted data pages from DRAM in SSD and therefore, increases the number of writes in SSDs.
For instance, in \emph{Stg\_1} workload, TICA-WED reduces the SSDs lifetime by more than 7x.
TICA-A which tries to balance the performance and endurance has only 4.7\% on 
average lifetime overhead compared to uCache {and S-RAC}.
Since TICA employs an unbalanced writing scheme between WO-SSD and RO-SSD, the lifetime of the RO-SSD is not affected by the increase in the number of writes.
Note that TICA-A still improves the SSDs lifetime by an average of 38\% 
compared to the single-level SSD architectures (not shown in Fig. 
\ref{fig:writes}).

%In addition to performance and power consumption enhancements, TICA reduces the number of writes in SSDs and therefore improves the SSD lifetime.
%In the traditional caching architectures, a read miss results in a write in SSD which shortens the SSD lifetime even in read-intensive workloads.
%TICA moves the read miss data pages to DRAM and prevents this type of writes to be committed to SSDs.
%Fig. \ref{fig:writes} shows the normalized number of writes compared to the traditional caching architecture.
%The lifetime improvement is 57\% on average and up to 93\% in the \emph{stg} workload.
%Since TICA has no write penalty and lower performance penalty in case of read miss compared to the traditional architecture, it provides superior performance and lifetime even in workloads with low hit ratio (such as \emph{stg} as reported in Fig. \ref{fig:writes}).
%Since TICA asynchronously moves dirty data pages from DRAM to RO-SSD, frequently written data pages such as filesystem metadata
%blocks can be hit again in DRAM before flushing to RO-SSD which reduces number of writes in RO-SSD.

\begin{figure*}[t]
%	\noindent\colorbox{amber}{
%		\parbox{\textwidth}
%		{
			\centering
			\includegraphics[scale=1.2]{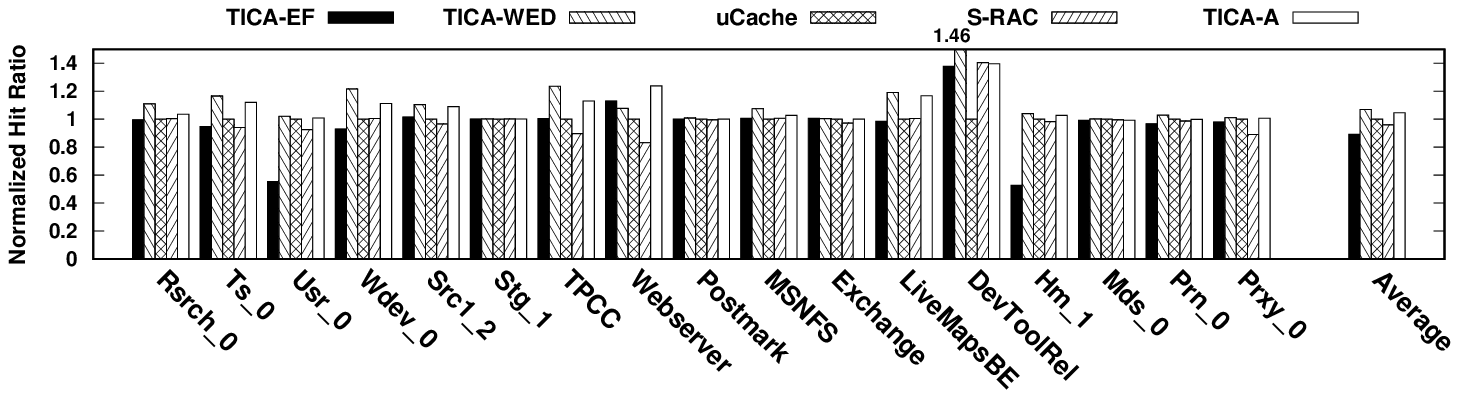}
			%\vspace{-0.3cm}
			\caption{Hit Ratio of Caching Architectures}
			\label{fig:hitratio}
			%\vspace{-0.6cm}
%		}
%	}
\end{figure*}

\subsection{Hit Ratio}
{TICA, uCache, and S-RAC} do not comply with a simple 
LRU algorithm (there is not a global LRU queue for all data pages).
Hence, the hit ratio of such multi-level caching architectures needs to be evaluated.
Although the performance of such architectures is already investigated in Section \ref{sec:performance}, the hit ratio evaluation enables us
to predict the performance in other hardware setups such as using RAID or \emph{Storage Area Network} (SAN) storages as the backend of the caching architecture.

Fig. \ref{fig:hitratio} shows the normalized hit ratio of TICA-EF, TICA-WED, 
{TICA-A, and S-RAC} compared to uCache.
TICA-WED which is optimized toward performance, has {the} highest average 
hit ratio.
TICA-A improves hit ratio compared to uCache {and S-RAC} in all workloads by 
an average of {6.4\%}.
The highest improvement belongs to \emph{DevToolRel} workload where TICA-WED 
and TICA-A improve hit ratio by 46\% and 39\%, respectively.
{S-RAC, however, has comparable hit ratio with TICA in this workload.
This is due to the ghost queue employed in S-RAC to identify requests with 
higher benefit for caching.}
Due to the DRAM eviction policy of TICA-EF, it has lower hit ratio compared to uCache and other TICA policies.
The hit ratio degradation of TICA-EF, however, is negligible in most workloads.

%The proposed architecture does not comply with a simple LRU algorithm (there is no global LRU queue for all data pages in TICA) unlike traditional architectures.
%Therefore, TICA hit ratio might differ from baseline LRU.
%Fig. \ref{fig:hitratio} compares the hit ratio of the proposed and traditional architectures.
%With the exception of \emph{usr} and \emph{MSNFS} workloads, TICA has a close hit ratio to the traditional architecture and it also provides a slightly higher hit ratio in 33\% of workloads.
%The average reduction in the hit ratio over all workloads is 3\%.

\subsection{Reliability}
{uCache and S-RAC employ} two WO-SSDs while TICA uses a WO-SSD alongside a 
RO-SSD in its architecture.
{Both uCache and S-RAC} will fail only when both WO-SSDs are failed.
There are two conditions which can result in failure of TICA: a) failure of WO-SSD and RO-SSD, and b) failure of WO-SSD and DRAM.
Since RO-SSD has lower \emph{Mean Time To Failure} (MTTF) {\cite{mttf}} compared to WO-SSD, TICA might reduce the overall reliability of the system.
DRAM, on the other hand, has higher MTTF compared to WO-SSD.
Therefore, the actual reliability of TICA depends on the duration of keeping dirty data pages in DRAM and RO-SSD.

{The reliability of caching architectures is calculated based on 
\emph{Reliability\space Block\space Diagram} (RBD) 
\cite{Dubrova:2013:FD:2462571}.
To calculate the system reliability, RBD uses 1) the reliability of system 
components (storage devices in our use-case) and 2) the dependency of system 
failure to the failure of components.
The reliability of storage devices is computed based on MTTF \cite{mttf}.
This is done via considering the exponential distribution for faults in SSDs, 
which is formulated in Equation \ref{eq:reliability}.
The MTTF value for storage devices is extracted from their datasheets.
Although other distributions such as Weibull might be more suitable, they 
require additional parameters to MTTF to model reliability. 
Field studies in SSD failure models do not disclose the brands of SSDs 
\cite{Narayanan:2016:SFD:2928275.2928278,Meza:2015:LSF:2745844.2745848,194414}, 
and therefore, we cannot use such models.
If such field studies become available, we can employ a more accurate MTTF for 
devices in the exponential distribution.
This can be done by estimating the real MTTF of the device, based on the} 
\emph{Cumulative Distribution Function} {(CDF) of 
Weibull distribution, as discussed in \cite{8412531}.}
The description of parameters in Equation \ref{eq:reliability} is available in Table \ref{tbl:paramdesc}.
Equation \ref{eq:ticareliability} and Equation \ref{eq:ucachereliability} show 
the formula for calculating the reliability of TICA and uCache, respectively.
{Note that the reliability of S-RAC is calculated using the same formula as 
uCache.}
The $\alpha$ variable denotes the weight of each failure scenario.
In the traditional RAID architectures, $\alpha$ is equal to one.
In TICA, $\alpha$ depends on the running workload and number of write requests.
Since TICA employs a RO-SSD instead of WO-SSD, compared to uCache {and 
S-RAC}, it is expected that TICA slightly reduces the reliability.
Considering 0.8 as the value of $\alpha$, which is close to the actual value of 
$\alpha$ in our experiments, TICA will have unreliability of $1.27*10^{-5}$ 
while unreliability of uCache {and S-RAC} is $1.14*10^{-5}$.
Note that the cost of hardware in TICA is lower than uCache and TICA will have the same reliability compared to uCache if the same hardware is employed for both architectures.

\begin{figure}[h]
	\scriptsize
	\vspace{-0.4cm}
	\begin{flalign}
	\hspace{-1cm}
	\label{eq:reliability}
	R_{Device}=&e^{-\frac{1}{MTTF_{Device}*365*24}} \\
	\label{eq:ticareliability}
	R_{TICA}=&\alpha*(1-(1-R_{\text{WO-SSD}})*(1-R_{D}))+\nonumber \\
	&(1-\alpha)*(1-(1-R_{\text{WO-SSD}})*(1-R_{\text{RO-SSD}})) \\
	\label{eq:ucachereliability}
	R_{uCache}=&1-(1-R_{\text{WO-SSD}})*(1-R_{\text{WO-SSD}})
	\end{flalign}
	\vspace{-0.4cm}
\end{figure}

%Baseline architectures in the experiments consist of two SSDs in a RAID 1 configuration.
%Data loss occurs when both SSDs fail at the same time and the size of lost data will be equal to the size of SSDs.
%Clean data pages in cache, however, are not lost since they still exist in HDD.
%Therefore, the amount of dirty data pages in cache should be considered while computing BLPY.
%The mentioned scenarios for data loss in TICA should be considered separately since they have different size of lost data.
%Fig. \ref{fig:reliability} depicts the BLPY for TICA and the baseline architectures.
%TICA has slightly higher reliability since in case of both SSD failures, a portion of dirty data pages exists in DRAM and vice versa.

\subsection{Overall}
\begin{figure}[t]
%	\noindent\colorbox{amber}{
%	\parbox{.48\textwidth}
%	{
	\centering
	\includegraphics[scale=0.7]{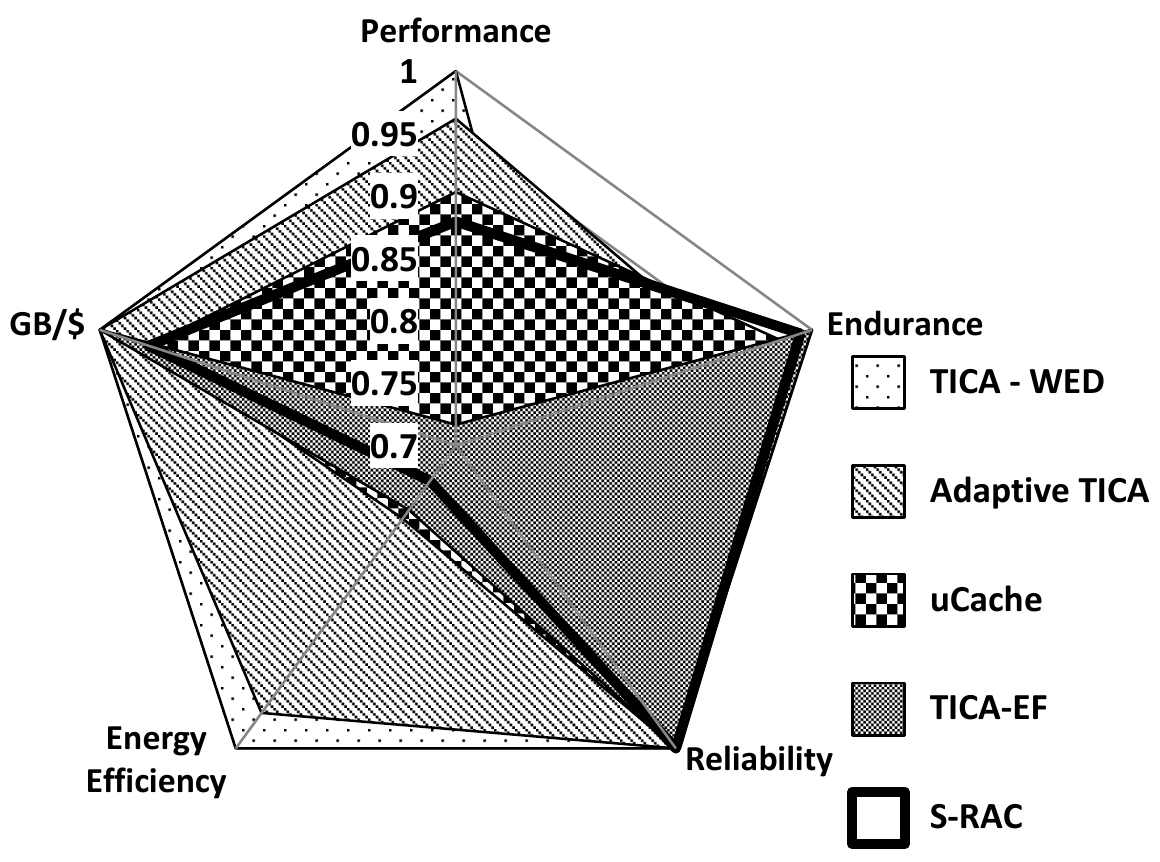}
	%\vspace{-0.3cm}
	\caption{Overall Comparison of Caching Architectures (Higher values are better)}
	\label{fig:overall}
	%\vspace{-0.6cm}
%}
%}
\end{figure}

\begin{figure}[t]
%	\noindent\colorbox{amber}{
%		\parbox{.48\textwidth}
%		{
			\centering
			\includegraphics[scale=0.65]{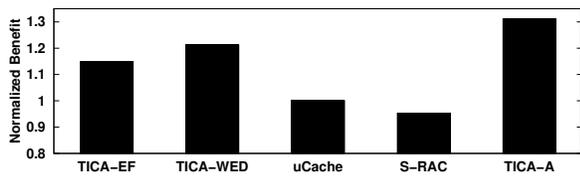}
			%\vspace{-0.3cm}
			\caption{Normalized Benefit of Various Caching Architectures}
			\label{fig:benefit}
			%\vspace{-0.6cm}
%		}
%	}
\end{figure}

We can conclude that the experimental results with following observations:
1) TICA improves performance and hit ratio compared to previous state-of-the-art architectures.
2) The power consumption is also improved in TICA by reducing the number of accesses to the SSDs.
3) Lifetime of SSDs is extended in TICA compared to single-level SSD caching 
architectures while the lifetime is negligibly reduced compared to uCache 
{and S-RAC}.
4) The reliability of TICA is the same as previous studies when the same hardware is employed. Reducing the total cost in TICA can result in slightly less reliability.
Fig \ref{fig:overall} shows the overall comparison of TICA policies with uCache 
{and S-RAC}.
All parameters are normalized to the highest value where higher values are better in all parameters.
{Fig. \ref{fig:benefit} also shows the overall benefit of caching 
architectures. Benefit is computed by multiplying normalized performance, 
endurance, cost, and power consumption. uCache and S-RAC, which focus on 
optimizing only one parameter have lower benefit compared to TICA variants. 
TICA-A provides the highest benefit since it considers all mentioned parameters 
in designing caching architecture and balances the performance and endurance, 
based on the workload characteristics.}

%TICA improves performance, power consumption, endurance, and reliability of caching architectures while having minimal cost overhead.
%Fig \ref{fig:overall} shows the overall comparison of TICA with the baseline architectures.
%All parameters are normalized to TICA where higher values are better in all parameters.
%Endurance and performance are two parameters with the highest improvement in TICA while endurance and power efficiency are also improved slightly.

\section{Conclusion}
\label{sec:conclusion}
In this paper, we demonstrated that simultaneously employing different SSDs in traditional architectures is not performance-efficient.
In addition, state-of-the-art architectures neglected to consider all aspects of the caching architectures.
To mitigate such problems, we proposed a three-level caching architecture, called TICA, which by employing RO-SSD and WO-SSD tries to reduce the cost and improve the performance and power consumption.
TICA does not have any single point of failure offering high reliable I/O cache architecture.
This is while the endurance cost of the proposed architecture is only {4.7\% 
}
higher than state-of-the-art caching architectures.
Additionally, the hardware cost of TICA is {5\%} less than conventional 
architectures.
The SSDs lifetime is extended by up to {38\%} compared to single-level SSD 
caching architectures.
The experimental results demonstrated that our architecture can improve 
performance and power consumption compared to previous studies, by up to 
{8\%} and {28\%}, respectively.

%This paper proposed a caching architecture, called TICA, that employs DRAM and read- and write-optimized SSDs in a three-level hierarchy in order to improve the performance and lifetime of the I/O cache.
%In addition to significant performance and lifetime improvement, TICA has no single point of failure offering high reliable I/O cache architecture. 
%This is while the cost penalty of the proposed architecture is only 15\% higher than the traditional caching architectures.
%The experimental results demonstrated that our architecture can improve performance, power consumption, and lifetime by 31\%, 11\%. and 57\%, respectively.

%To furthermore improve the proposed architecture, we plan to follow these steps in the future work: 
%a) the reliability of the proposed architecture can be evaluated accurately by calculating the total dirty pages residing on each storage device and by computing the \emph{Mean Time To Failure} (MTTF) and \emph{Uncorrectable Error Rate} (UER) of devices.
%b) Improving the eviction policy of DRAM by evicting the data pages with high probability of future references to SSDs instead of discarding them from cache.
%c) Analyzing the cache eviction policies proposed in previous studies and selecting the best policy for each device and customizing it towards higher performance and lower write count.

%\begin{figure}[t]
%	\centering
%	\includegraphics[scale=0.55]{reliability.eps}
%	%\vspace{-0.3cm}
%	\caption{Normalized BLPY of Caching Architectures}
%	\label{fig:reliability}
%	%\vspace{-0.6cm}
%\end{figure}

\section*{Acknowledgments}
This work has been partially supported by \emph{Iran National Science Foundation} (INSF) under grant number 96006071 and by HPDS Corp.

\bibliographystyle{IEEEtran}
\bibliography{IEEEabrv,ref}
\begin{IEEEbiography}[{\includegraphics[width=1in,height=1.25in,clip,keepaspectratio]{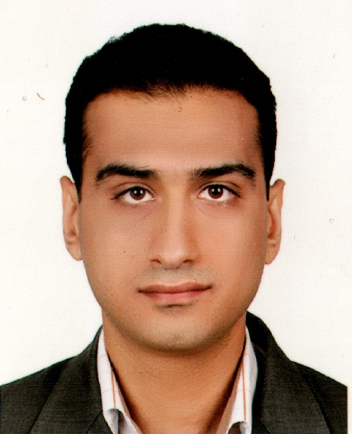}}]{Reza Salkhordeh}
	received the B.Sc. degree in computer engineering from Ferdowsi University of Mashhad in 2011, and M.Sc. degree in computer engineering from Sharif University of Technology (SUT) in 2013. He has been a member of \emph{Data Storage, Networks, and Processing} (DSN) lab since 2011. He was also a member of Iran National Elites Foundation from 2012 to 2015. He has been the director of Software division in HPDS corporation since 2015. He is currently a Ph.D. candidate at SUT. His research interests include operating systems, solid-state drives, memory systems, and data storage systems.
\end{IEEEbiography}
\vskip -3.5\baselineskip plus -1fil
\vfill
%\vspace{-.6cm}
\begin{IEEEbiography}[{\includegraphics[width=1in,height=1.25in,clip,keepaspectratio]{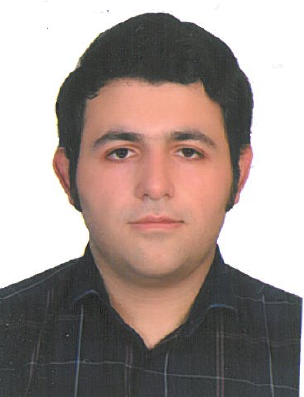}}]{Mostafa Hadizadeh}
received the B.Sc. degree in computer engineering from Shahid Beheshti University (SBU), Tehran, Iran, in 2016. He is currently pursuing the M.Sc. degree in computer engineering at Sharif University of Technology (SUT), Tehran, Iran. He is a member of Data Storage, Networks, and Processing (DSN) Laboratory from 2017. From December 2016 to May 2017, he was a member of Dependable Systems Laboratory (DSL) at SUT. His research interests include computer architecture, memory systems, dependable systems and systems on chip.
\end{IEEEbiography}
\vskip -3.5\baselineskip plus -1fil
\vfill
%\vspace{-.6cm}
\begin{IEEEbiography}[{\includegraphics[width=1in,height=1.25in,clip,keepaspectratio]{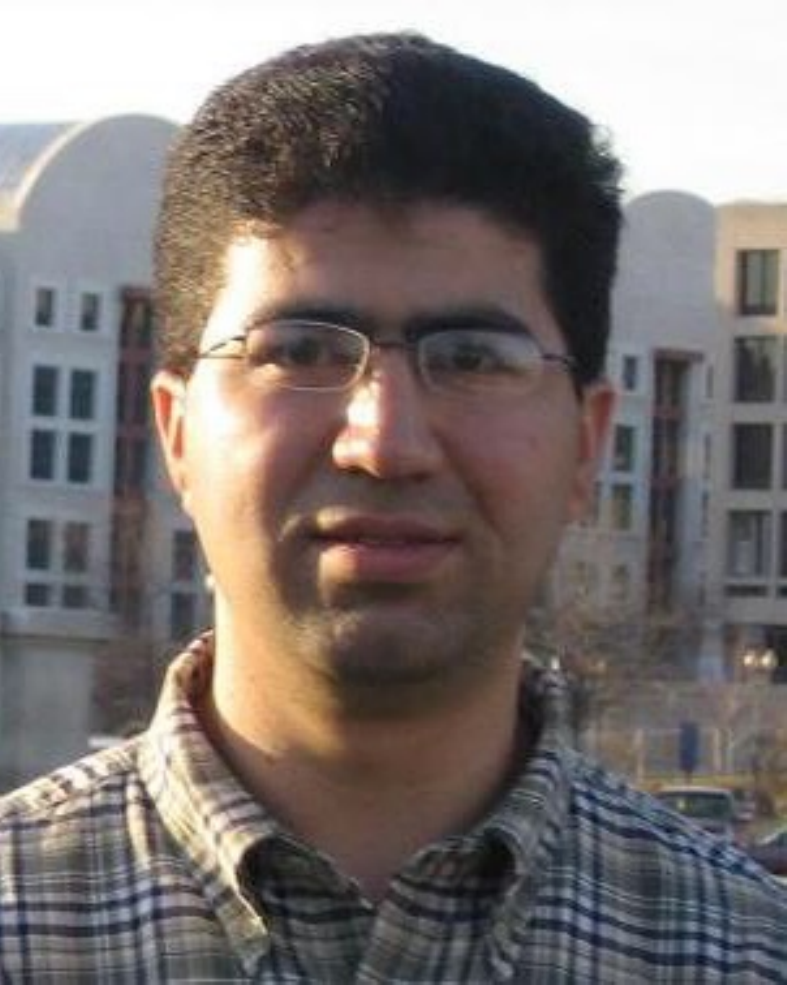}}]{Hossein Asadi}
	(M'08, SM'14) received the B.Sc. and M.Sc. degrees in computer engineering 
	from the SUT, Tehran, Iran, in 2000 and 2002, respectively, and the Ph.D. 
	degree in electrical and computer engineering from Northeastern University, 
	Boston, MA, USA, in 2007. 
	
	He was with EMC Corporation, Hopkinton, MA, USA, as a Research Scientist 
	and Senior Hardware Engineer, from 2006 to 2009. From 2002 to 2003, he was 
	a member of the Dependable Systems Laboratory, SUT, where he researched 
	hardware verification techniques. From 2001 to 2002, he was a member of the 
	Sharif Rescue Robots Group. He has been with the Department of Computer 
	Engineering, SUT, since 2009, where he is currently a tenured Associate 
	Professor. He is the Founder and Director of the \emph{Data Storage, 
	Networks, and Processing} (DSN) Laboratory, Director of Sharif 
	\emph{High-Performance Computing} (HPC) Center, the Director of Sharif 
	\emph{Information and Communications Technology Center} (ICTC), and the 
	President of Sharif ICT Innovation Center. He spent three months in the 
	summer 2015 as a Visiting Professor at the School of Computer and 
	Communication Sciences at the Ecole Poly-technique Federele de Lausanne 
	(EPFL). He is also the co-founder of HPDS corp., designing and fabricating 
	midrange and high-end data storage systems. He has authored and co-authored 
	more than eighty technical papers in reputed journals and conference 
	proceedings. His current research interests include data storage systems 
	and networks, solid-state drives, operating system support for I/O and 
	memory management, and reconfigurable and dependable computing.
	
	Dr. Asadi was a recipient of the Technical Award for the Best Robot Design 
	from the International RoboCup Rescue Competition, organized by AAAI and 
	RoboCup, a recipient of Best Paper Award at the 15th CSI International 
	Symposium on \emph{Computer Architecture \& Digital Systems} (CADS), the 
	Distinguished Lecturer Award from SUT in 2010, the Distinguished Researcher 
	Award and the Distinguished Research Institute Award from SUT in 2016, and 
	the Distinguished Technology Award from SUT in 2017. He is also recipient 
	of Extraordinary Ability in Science visa from US Citizenship and 
	Immigration Services in 2008. He has also served as the publication chair 
	of several national and international conferences including CNDS2013, 
	AISP2013, and CSSE2013 during the past four years. Most recently, he has 
	served as a Guest Editor of IEEE Transactions on Computers, an Associate 
	Editor of Microelectronics Reliability, a Program Co-Chair of CADS2015, and 
	the Program Chair of CSI National Computer Conference (CSICC2017). 
\end{IEEEbiography}

\end{document}